\documentclass[a4paper,10pt,reqno]{amsart}
\usepackage[numbers]{natbib}

\usepackage{stmaryrd}
\usepackage{amscd}
\usepackage{bm}
\usepackage{amssymb}
\usepackage{amsmath}
\usepackage{pgfplots}
\usepackage{subfig}
\usepackage{graphicx}

\newcommand{\R}{\ensuremath{\mathbb{R}}}
\newcommand{\Div}[2][]{\ensuremath{\,\text{div}_{#1}(#2)}}
\newcommand{\cof}[1]{\ensuremath{\,\text{cof}(#1)}}
\newcommand{\T}{\ensuremath{\mathcal{T}}}
\newcommand{\F}{\ensuremath{\mathcal{F}}}
\newcommand{\Jbnd}{\ensuremath{J_{b}}}
\newcommand{\tr}[1]{\ensuremath{\,\text{tr}(#1)}}
\newcommand{\re}[1]{\ensuremath{\textcolor{black}{\hat{#1}}}}
\newcommand{\ph}[1]{\ensuremath{\textcolor{black}{#1}}}
\newcommand{\HDiv}[1][]{\ensuremath{H(\text{div},#1)}}
\newcommand{\Hone}[1][]{\ensuremath{H^1}}
\newcommand{\Ltwo}[1][]{\ensuremath{L^2(#1)}}
\newcommand{\idop}{\ensuremath{\text{id}}}

\newcommand{\ddt}[1]{\ensuremath{\frac{\partial #1}{\partial t}}}

\newcommand{\Hphys}{\ensuremath{\ph{h}}}
\newcommand{\Href}{\ensuremath{\re{h}}}

\newcommand{\Sphys}{\ensuremath{\ph{S}}}

\newcommand{\sphys}{\ensuremath{\ph{s}}}
\newcommand{\sref}{\ensuremath{\re{s}}}

\newcommand{\nphys}{\ensuremath{\ph{n}}}
\newcommand{\nref}{\ensuremath{\re{n}}}

\newcommand{\tphys}{\ensuremath{\ph{\tau}}}
\newcommand{\tref}{\ensuremath{\re{\tau}}}

\newcommand{\fphys}{\ensuremath{\ph{f}}}
\newcommand{\fref}{\ensuremath{\re{f}}}

\newcommand{\xphys}{\ensuremath{\ph{x}}}
\newcommand{\xref}{\ensuremath{\re{x}}}

\newcommand{\uphys}{\ensuremath{\ph{u}}}
\newcommand{\uref}{\ensuremath{\re{u}}}

\newcommand{\vphys}{\ensuremath{\ph{v}}}
\newcommand{\vref}{\ensuremath{\re{v}}}

\newcommand{\pref}{\ensuremath{\re{p}}}

\newcommand{\lphys}{\ensuremath{\ph{\lambda}}}
\newcommand{\lref}{\ensuremath{\re{\lambda}}}

\newcommand{\Psiphys}{\ensuremath{\ph{\Psi}}}
\newcommand{\Psiref}{\ensuremath{\re{\Psi}}}

\newcommand{\Ophys}{\ensuremath{\ph{\Omega}_t}}
\newcommand{\Oref}{\ensuremath{\re{\Omega}}}

\newcommand{\Tref}{\ensuremath{\re{T}}}

\newcommand{\Gphys}{\ensuremath{\ph{\Gamma}_t}}
\newcommand{\Gref}{\ensuremath{\re{\Gamma}}}

\newcommand{\gradphys}{\ensuremath{\nabla_{\xphys}}}
\newcommand{\gradref}{\ensuremath{\nabla_{\xref}}}

\newcommand{\Fmat}{\ensuremath{\bm{F}}}
\newcommand{\Imat}{\ensuremath{\bm{I}}}
\newcommand{\Cmat}{\ensuremath{\bm{C}}}
\newcommand{\Emat}{\ensuremath{\bm{E}}}
\newcommand{\Pmat}{\ensuremath{\bm{P}}}
\newcommand{\stress}{\ensuremath{\bm{\sigma}}}
\newcommand{\Stress}{\ensuremath{\bm{\Sigma}}}
\newcommand{\hesse}{\ensuremath{\bm{H}}}

\usepackage{chngcntr}
\counterwithin{figure}{section}
\counterwithin{table}{section}
\counterwithin{equation}{section}

\newbox{\bigpicturebox}

\title[FSI with $H(\text{div})$-conforming finite elements]{Fluid-structure interaction with $H(\text{div})$-conforming finite elements}
\date{\today}

\author{Michael Neunteufel} 
\address[Michael Neunteufel]{Institute for Analysis and Scientific Computing, TU Wien, Wiedner Hauptstrasse 8-10, 1040 Wien, Austria.}
\email{michael.neunteufel@tuwien.ac.at}
\urladdr{https://www.asc.tuwien.ac.at/~schoeberl/wiki/index.php/Michael\_Neunteufel}

\author{Joachim Sch\"oberl} 
\address[Joachim Sch\"oberl]{Institute for Analysis and Scientific Computing, TU Wien, Wiedner Hauptstrasse 8-10, 1040 Wien, Austria.}
\email{joachim.schoeberl@tuwien.ac.at}
\urladdr{https://www.asc.tuwien.ac.at/~schoeberl/wiki/index.php/Joachim\_Sch{\"o}berl}

\begin{document} 
\maketitle

\begin{abstract}
In this paper a novel application of the (high-order) $H(\text{div})$-conforming Hybrid Discontinuous Galerkin finite element method for monolithic fluid-structure interaction (FSI) is presented. The Arbitrary Lagrangian Eulerian (ALE) description is derived for $H(\text{div})$-conforming finite elements including the Piola transformation, yielding exact divergence free fluid velocity solutions. The arising method is demonstrated by means of the benchmark problems proposed by Turek and Hron \cite{TH06}. With hp-refinement strategies singularities and boundary layers are overcome leading to optimal spatial convergence rates.\\

\end{abstract}

\section{Introduction}
\label{sec:intro}
Fluid-structure interaction plays a crucial role in physics, where fluids interact with elastic, solid structures and affect each other. Such problems arise in a wide variety in nature and technology, e.g., in the simulation of blood vessels \cite{PESKIN77,FQV10,CPP13}, in material processes, or in aerodynamics \cite{BHKWB11,CCRTT19}, to name only a few. The Tacoma Narrow bridge is an infamous example of the importance of FSI.\newline
The coupling part in multiphysics problems is still challenging and a vast amount of research has been invested in finding a stable and efficient discretization scheme. For the spatial discretization of the (incompressible) Navier--Stokes equations the famous $P2$--$P1$ Taylor--Hood elements \cite{HT73} are widely used, yielding $\Hone$-conforming low order methods. In the last two decades high-order methods for fluid-structure interaction were investigated \cite{FP14,CPP13,PP10,PPB07}. 

Discontinuous Galerkin (DG) methods for Navier--Stokes were developed in \cite{Cockburn05,Cockburn07}, entailing beneficial stability and conservation properties. To avoid the disadvantage of strong coupling between elements and the high number of degrees of freedom, Hybrid Discontinuous Galerkin (HDG) methods have been introduced in the context of mixed finite elements \cite{BF91}, successfully developed for the Navier--Stokes equations \cite{Cesmelioglu2013,RC12}, and recently applied to FSI \cite{SSJ16}. Using $H^1$-conforming Taylor--Hood elements have the significant drawback of velocity fields not being exactly divergence free, i.e., $\Div{u}\not\equiv0$ point-wise. Therefore, exact incompressible finite element methods were developed \cite{CCS06,CG05,CG05_2,RW18,KS12,GLS19,SL18} including  $H(\text{div})$-conforming elements. Instead of decoupling the elements completely as suggested in HDG techniques, the $H(\text{div})$-conforming HDG method introduced in \cite{Lehrenfeld10,LS15} does not break the normal continuity between elements. A second approach yielding robust methods enforces weakly the divergence-free constraint and inter-element continuity of the normal velocity by stabilization terms \cite{JDSST16,KNWK17}.

Fluid problems are commonly given in Eulerian form, whereas in solid elasticity the equations are formulated mostly in Lagrangian form. To combine both approaches and equations the Arbitrary Lagrangian Eulerian form was developed and has been discussed intensively in the finite element context of $\Hone$-conforming finite elements \cite{DGH82,DH04,Donea2003,HAC94}. 
Recently, an ALE-DG formulation has been proposed in \cite{FHWK20}, including the weakly divergence-free constraints, and in the space-time setting $H(\text{div})$-conforming elements have been considered on moving domains \cite{HS19}. However, using $H(\text{div})$-conforming elements in the context of ALE, an adaption of the form is needed due to the different transformation rules for these elements, namely the Piola transformation. The latter results in two additional terms supplementing the classical mesh-velocity. One main contribution of this work is to derive the $H(\text{div})$ based ALE transformation and embed it in terms of FSI enabling exact divergence free velocity solutions in the fluid domain.

Focusing on the fluid part, the elastic wave equation will be discretized with Lagrangian finite elements. Using two different finite element spaces for the velocity necessitates the use of Lagrange multipliers to couple the solid with the fluid.\newline

This paper is structured as follows. In the next section standard notation including the Navier--Stokes and elastic wave equations are introduced. In Section 3 properties of the $H(\text{div})$-conforming elements are described and the corresponding novel ALE formulation is derived. In Section 4 a spatial monolithic formulation for the fluid-structure interaction problem based on the $H(\text{div})$-conforming HDG Navier--Stokes equations is introduced and the time discretization scheme is discussed. Numerical examples are given in the last section, confirming the efficiency of the presented method. Therein, the arising singularities and boundary layers are resolved using hp-refinement strategies.

\section{Preliminaries and equations}
\label{sec:prel}
\subsection{Notation}
\label{subsec:notation}
We assume a bounded domain $\Ophys\subset \R^d$, with $d\in\{2,3\}$ and a smooth boundary $\partial\Ophys$, which can move in time $t\in[0,T]$ and is divided into a fluid and a solid domain, $\Ophys^f$ and $\Ophys^s$, respectively. The interface on which the different domains interact is given by $\Gphys=\overline{\Ophys^f}\cap\overline{\Ophys^s}$. Furthermore, we define the initial configurations $\Oref^s:=\ph{\Omega}^s_0$, $\Oref^f:=\ph{\Omega}^f_0$ as the reference domain and $\Gref:=\ph{\Gamma}_0$ as the reference interface.

We denote by $\langle\cdot,\cdot\rangle_{\Omega}$ and $\langle\cdot,\cdot\rangle_{\partial\Omega}$ the $L^2$-inner product over a domain $\Omega$ and  over a boundary $\partial\Omega$, respectively. For the Euclidean norm $\|\cdot\|_2$ we will neglect the subscript. 

In the discretized setting we assume a  shape regular finite element mesh $\T_h$ of the domain $\Omega$ consisting of (possibly curved) triangles and quadrilaterals in 2D or tetrahedral, prism, hexahedron, and pyramids in three space dimensions. The subscript $h$ indicates discretized objects if not specified otherwise. The set of all interfaces between two elements, edges in 2D and faces in 3D, respectively, are called facets, which we will denote by $\F_h$. The set of all piece-wise polynomials up to degree $k$ on the triangulation $\T_h$ and the skeleton $\F_h$ is given by $\Pi^k(\T_h)$ and $\Pi^k(\F_h)$, respectively.

Due to the huge number of different test and trial functions, we will denote all test functions by $\Psi$ and, if necessary, add a superscript referring to the corresponding unknowns, e.g., the test function to the unknown $p$ is given by $\Psi^{p}$.

\subsection{Equations}
\label{sec:equations}
\subsubsection{Fluid}
\label{subsubsec:fluid}
On the fluid domain the incompressible, unsteady, Newtonian Navier--Stokes equations are solved, which are given in Eulerian form
\begin{align}
&\rho^f\ddt{v^f}+\rho^f(v^f\cdot\nabla)v^f-\Div{\stress^f}=f& \text{on }\Ophys^f,\nonumber\\
&\Div{v^f} = 0& \text{on }\Ophys^f,
\end{align}
where the fluid stress tensor $\stress^f$ is
\begin{align}
\stress^f = -p^f\Imat+2\rho^f\nu^f\varepsilon(v^f)
\end{align}
and $\Imat$ denotes the identity matrix. The fluid velocity is denoted by $v^f$ and the pressure by $p^f$. The parameters are the fluid density $\rho^f$ and the kinematic viscosity $\nu^f$. The symmetric part of the gradient is given by the function
\begin{align}
\varepsilon(v):=\frac{1}{2}((\nabla v)^T+\nabla v),
\end{align}
where the gradient of a function is defined as
\begin{align}
\nabla v := \left(\frac{\partial v_i}{\partial x_j}\right)_{i,j=1}^d.
\end{align}
\subsubsection{Solid}
\label{subsubsec:solid}
The elastic wave equation in Eulerian form reads
\begin{align}
 \rho^s\frac{\partial^2 u^s}{\partial t^2}+\rho^s\nabla v^sv^s-\Div{\stress^s}=g\quad\text{ on }\Ophys^s\label{eq:el_wave_eul}
\end{align}
and in Lagrangian form
\begin{align}
\rho^s\frac{\partial^2 u^s}{\partial t^2}-\Div{\Pmat^s}=g\quad\text{ on }\Oref^s.\label{eq:el_wave}
\end{align}
Here, $\stress^s$ denotes the Cauchy stress tensor, $u^s$ the solid displacement and $\rho^s$ the solid density. The first Piola--Kirchhoff stress tensor $\Pmat^s:=J\stress^s\Fmat^{-T}$ can be expressed by the deformation gradient, its determinant
\begin{align}
\Fmat=\Imat+\nabla u^s,\qquad J=\det(\Fmat),
\end{align}
and the second Piola--Kirchhoff stress tensor $\Stress^s$
\begin{align}
\Pmat^s=\Fmat\Stress^s.
\end{align}
The solid velocity $v^s$ is defined by $v^s=\ddt{u^s}$. In the following we will use the material law of St. Venant--Kirchhoff
\begin{align}
\Stress^s =\lambda^s\tr{\Emat}\Imat+2\mu^s\Emat,
\end{align}
with the Green strain tensor 
\begin{align}
\Emat:=\frac{1}{2}(\Cmat-\Imat),\quad \Cmat:=\Fmat^T\Fmat,
\end{align}
where $\Cmat$ denotes the Cauchy--Green strain tensor. The two material parameters $\lambda^s$ and $\mu^s$ are the Lam{\'e} coefficients, which can be computed with $E^s$ and $\nu^s$, the Young's modulus and the Poisson's ratio, respectively
\begin{align}
&\nu^s = \frac{\lambda^s}{2(\lambda^s+\mu^s)}, \quad\quad\quad\quad\quad E^s = \frac{\mu^s(3\lambda^s+2\mu^s)}{\lambda^s+\mu^s},\nonumber\\
&\lambda^s =\frac{E^s\nu^s}{(1+\nu^s)(1-2\nu^s)},\quad\quad\, \mu^s =\frac{E^s}{2(1+\nu^s)}.
\end{align}

\subsection{Interface and boundary conditions}
\label{subsec:int_bnd_cond}
To obtain a correct coupling behavior we have to enforce continuity of the fluid and solid velocity over the interface and that the forces are in equilibrium
\begin{align}
v^s=v^f,\quad\quad \stress^sn^s=\stress^fn^f,\quad\text{ on }\Gphys,
\end{align}
where $n^f$ and $n^s$ denote the fluid and solid outer normal vector, respectively, on the interface $\Gphys$. In terms of the fluid part, this can be seen as no-slip condition on the interface. On the other boundaries we prescribe the standard Dirichlet and Neumann boundary conditions for the fluid and solid
\begin{align}
v^f=v_D,\,\, \stress^fn = f_N,\,\, u^s=u_D,\,\, \stress^sn=g_N.
\end{align}

\section{ALE for $H(\text{div})$-conforming methods}
\label{sec:ALE_NavStokes}
As the Navier--Stokes equations are given in Eulerian and the elastic wave equation in Lagrangian form, the ALE description is used to transform the Navier--Stokes equations from the current configuration $\Ophys$ to the reference domain $\Oref$. Another approach would be to transform the elastic wave equation into its Eulerian form and use a pure Eulerian description \cite{RW10,Richter2013,FRW16}. This leads to an additional convection term appearing in the elastic wave equation and the system can be interpreted as a two-phase problem. Also XFEM based methods on fixed grids \cite{GW08,WG06} have been introduced avoiding remeshing and recently a CutFEM based method has been proposed \cite{SAW19}. In this work, however, we will not consider this approaches and use the ALE description.

For the Readers convenience we first give a short revision of the standard ALE description form. Then the $H(\text{div})$-conforming finite element spaces are introduced and the ALE form together with the Piola transformation is discussed.

\subsection{ALE for $\Hone$-conforming methods}
\label{subsec:ALE_stand}
For the Arbitrary Lagrangian Eulerian description we assume a time dependent, invertible and sufficiently smooth function $\Phi$ between the reference and spatial domain
\begin{align}
& \Phi: \Oref\times[0,T]\rightarrow\Ophys\times[0,T],\nonumber\\
& (\xref,t)\mapsto \Phi(\xref,t)=(\varphi(\xref,t),t)=(\xphys,t),\label{eq:alefunc}
\end{align}
where $\varphi$ is called the deformation function.

A function $\fphys:\Ophys\times[0,T]\rightarrow\R^d$ is coupled with $\fref:\Oref\times[0,T]\rightarrow\R^d$ via the relation
\begin{align}
&\fphys\circ\Phi=\fref\label{eq:rel_spat_ref}.
\end{align}

Differentiating \eqref{eq:rel_spat_ref} with respect to time and space in reference coordinates $\xref$ and using the chain rule yields the following transformation rules
\begin{align}
&\gradphys \fphys\circ\Phi \gradref\varphi=\gradref\fref,\label{eq:trafo_grad}\\
&\ddt{\fphys}\circ\Phi = \ddt{\fref}-\gradref\fref \Fmat^{-1}\ddt{\varphi},\label{eq:trafo_time}
\end{align}
where $\gradphys$ and $\gradref$ denote the gradients with respect to spatial or reference coordinates $\xphys$ or $\xref$, respectively. The gradient of $\varphi$ is called the deformation gradient, which will be denoted in the following by $\Fmat:=\gradref\varphi$ and $J:=\det(\Fmat)$. The time derivative of the mesh deformation function $\varphi$ is called the mesh-velocity, describing the relative motion of the mesh and is defined in what follows by $\dot{\varphi}$.

The deformation function $\varphi$ is assumed to be in $[\Hone[\Oref]]^d$ and we define the Lagrange nodal finite element space $U_h$ for the deformation and displacement as
\begin{align}
U_h:=[\Pi^k(\T_h)]^d\cap C(\Oref,\R^d),
\end{align}
where $C(\Oref,\R^d)$ denotes the set of all vector valued continuous functions.

\subsection{H(\text{div})-conforming elements}
\label{subsec:HDiv_el}
The function space $\HDiv[\Omega]$ is defined as the space of all square integrable functions $[\Ltwo[\Omega]]^d$, where the weak divergence is also square integrable
\begin{align}
\HDiv[\Omega]:=\{u\in [\Ltwo[\Omega]]^d|\,\Div{u}\in \Ltwo[\Omega]\}.
\end{align}
To ensure that a function $\uref\in \HDiv[\Oref^f]$ is in the space $\HDiv[\Ophys^f]$ after deformation, the so-called Piola transformation is used
\begin{align}
\uphys\circ\Phi=P_{\Phi}[\uref]:=\frac{1}{J}\Fmat\uref.\label{eq:piola_trafo}
\end{align}
If the deformation $\Phi$ is obvious, we will neglect the subscript of the Piola transformation.\newline

Let $\Phi:\hat{T}\rightarrow T$ be a diffeomorphic mapping from the reference element $\hat{T}$ to the physical element $T$ and $\Psi$ a diffeomorphic mapping from $T$ to another physical element $\tilde{T}$. Let $\hat{\sigma}\in \HDiv[\hat{T}]$. Then, the Piola transformation \eqref{eq:piola_trafo} has the following well known properties \cite{boffi13, Raviart1977}:
\begin{enumerate}
\item $\sigma$ is in the space  $\HDiv[T]$ with
\begin{align}
\label{eq:piola_trafo_div}
\Div[\xphys]{\sigma}\circ\Phi=J^{-1}\Div[\xref]{\hat{\sigma}}.
\end{align}  
\item Let furthermore $\hat{e}$ be an edge of the reference element and $e=\Phi(\hat{e})$. Then 
\begin{align}
\langle\sigma, n_e\rangle_e=\langle\hat{\sigma},n_{\hat{e}}\rangle_{\hat{e}}.\label{eq:norm_flow_piola}
\end{align}
\item With $\Theta:=\Psi\circ\Phi$ there holds
\begin{align}
\label{eq:piola_trafo_comp}
 P_{\Psi\circ\Phi}[\hat{\sigma}]=P_{\Theta}[\hat{\sigma}]=P_{\Psi}[P_{\Phi}[\hat{\sigma}]].
\end{align}
\end{enumerate}

The $H(\text{div})$-conforming finite element spaces of Raviart--Thomas \cite{Raviart1977} and Brezzi--Douglas--Marini (BDM) \cite{Brezzi1985} fulfil the condition that the normal jump of the functions is zero over the elements. E.g., the BDM space of polynomial order $k$ is given by 
\begin{align}
W_h:=\{v_T\in[\Pi^k(\T_h)]^d\,|\,\llbracket v_T\cdot n\rrbracket_F= 0, \forall F\in \F_h\},\label{eq:hdiv_fespace}
\end{align}
where $\llbracket\cdot\rrbracket$ denotes the jump over elements.

The degrees of freedom are highly related to the normal flow through the faces. These are defined on a fixed reference element and due to \eqref{eq:norm_flow_piola} the normal continuity between elements is ensured also after the deformation on the physical element. For the construction of (high-order) $H(\text{div})$-conforming finite elements we refer to \cite{boffi13,Zaglmayr06}.

\subsection{ALE for H(\text{div})-conforming elements}
\label{subsec:ALE_HDIV_EL}
The connection of $H(\text{div})$-conforming functions between the spatial and reference configuration is given via the Piola transformation
\begin{align}
\fphys\circ\Phi=\frac{1}{J}\Fmat\fref,\qquad \Phi = (\varphi,\idop),
\end{align}
with $\varphi\in [H^1(\Oref^f)]^d$. Due to \eqref{eq:piola_trafo_comp} the composition of two Piola transformations is again a Piola transformation and thus, it is guaranteed that the function $\fphys$ is in $\HDiv[\Ophys^f]$.
As the Piola transformation itself depends on space and time, the derivatives \eqref{eq:trafo_grad} and \eqref{eq:trafo_time} need to be recalculated. Therefore, we first compute the derivatives of the Piola transformation. Note that in the discretized setting the deformation $\varphi$ and the $H(\text{div})$-conforming velocity $\vref_T$ are both piece-wise smooth functions on the triangles $\Tref$. Thus, we can compute the Hessian and the gradients on each triangle, see Appendix A for the computations.

Let $\Phi = (\varphi,\idop):\R^d\times[0,T]\rightarrow\R^d$ a piece-wise smooth deformation function on the triangulation $\T_h$, $\Fmat=\gradref\varphi$ and $J=\det(\Fmat)$. Then, with the notation $\hesse^i_{jk}:=\frac{\partial^2\varphi_i}{\partial \xref_j\xref_k}$ for the Hessian, there holds on each triangle $\Tref$
\begin{align}
\partial_{\xref_j}(\frac{1}{J}\Fmat u)_i&=-\frac{1}{J}\Fmat^{-T}:(\partial_{\xref_j}\Fmat)(\Fmat u)_i+\frac{1}{J}\left((\hesse^iu)_j+(\Fmat\nabla_{\xref} u)_{ij}\right),\\
\partial_t(\frac{1}{J}\Fmat u)&=\frac{1}{J}\left(\gradref\dot{\varphi}- \tr{\gradref\dot{\varphi}\Fmat^{-1}}\Fmat\right)u+\frac{1}{J}\Fmat\dot{u},
\end{align}
where $\tr{A}$ denotes the trace of $A$.

Hence, the ALE derivative transformations are given by
\begin{align}
\gradphys\fphys\circ\Phi &= \gradref (P_{\Phi}[\fref])\Fmat^{-1},\label{eq:trafo_grad_hdiv}\\
\ddt{\fphys}\circ\Phi &= \frac{1}{J}(\gradref\dot{\varphi}-\tr{\gradref\dot{\varphi}\Fmat^{-1}}\Fmat)\fref+P_{\Phi}[\ddt{\fref}] - \gradref P_{\Phi}[\fref]\Fmat^{-1}\dot{\varphi}.\label{eq:trafo_time_hdiv}
\end{align}

To obtain the ALE variational formulation for $H(\text{div})$-conforming elements we integrate $\ddt{\fphys}$ over the spatial domain $\Ophys^f$ and multiply with a test function $\Psiphys\in \HDiv[\Ophys^f]$. A change of variables, the transformation rules \eqref{eq:trafo_grad_hdiv} and \eqref{eq:trafo_time_hdiv} from above, and the Piola transformation $\Psiphys\circ\Phi=P_{\Phi}[\Psiref]$ with $\Psiref\in \HDiv[\Oref^f]$ for the test function yields
\begin{align}
\langle\ddt{\fphys},\Psiphys\rangle_{\Ophys} &= \langle J\ddt{\fphys}\circ\Phi,\Psiphys\circ\Phi\rangle_{\Oref}\nonumber\\
&=\langle J(\frac{1}{J}(\gradref\dot{\varphi}-\tr{\gradref\dot{\varphi}\Fmat^{-1}}\Fmat)\fref+P_{\Phi}[\ddt{\fref}]- \gradref P_{\Phi}[\fref]\Fmat^{-1}\dot{\varphi}),P_{\Phi}[\Psiref]\rangle_{\Oref}\nonumber\\
&=\langle J((\gradref\dot{\varphi}\Fmat^{-1}-\tr{\gradref\dot{\varphi}\Fmat^{-1}}\Imat)P_{\Phi}[\fref]+P_{\Phi}[\ddt{\fref}]- \gradref P_{\Phi}[\fref]\Fmat^{-1}\dot{\varphi}),P_{\Phi}[\Psiref]\rangle_{\Oref}.
\end{align}

In addition to the mesh velocity term
\begin{align}
-\gradref P_{\Phi}[\fref]\Fmat^{-1}\dot{\varphi}
\end{align}
from the standard ALE formulation, we obtain the additional terms
\begin{align}
\label{eq:add_term_ale_hdiv}
(\gradref\dot{\varphi}\Fmat^{-1}-\tr{\gradref\dot{\varphi}\Fmat^{-1}}\Imat)P_{\Phi}[\fref].
\end{align}

We note that one could have deduced the ALE formulation also in strong sense, which would have led to the same result.

\section{Discretization}
\subsection{Spatial discretization}
First, the new ALE description is used to transform the $H(\text{div})$-conforming Hybrid Discontinuous Galerkin method for the time dependent, incompressible Navier--Stokes equations from the spatial to the reference domain. Then the elastic wave equation is discretized with standard $\Hone$-conforming elements.

\subsubsection{H(\text{div})-conforming Hybrid Discontinuous Galerkin method for Navier--Stokes}
For $H(\text{div})$-conforming HDG the velocity is split into a normal and a tangential continuous part $v=(v_T, v_F)$, where $v_T\in W_h$, see \eqref{eq:hdiv_fespace}. The skeleton variable $v_F$ lives in the following facet space
\begin{align}
F_h :=\{v_F\in [\Pi^k(\F_h)]^d\,|\,v_F\cdot n=0\}.
\end{align}
Thus, the complete finite element space for the fluid velocity is defined as
\begin{align}
V_h:=W_h\times F_h.
\end{align}
The appropriate finite element space for the pressure is given by piece-wise polynomials of one polynomial degree less than the velocity space
\begin{align}
Q_h:=\Pi^{k-1}(\T_h).\label{eq:pressure_space}
\end{align}
With this choice of spaces there holds
\begin{align}
\label{eq:rel_divV_Q}
\Div{W_h}\subset Q_h,
\end{align}
which has the crucial consequence that from weak incompressibility there follows immediately strong incompressibility
\begin{align}
\langle\Div{v_T},\Psi\rangle_{\Oref} = 0,\, \forall \Psi \in Q_h \Longrightarrow \Div{v_T} = 0\text{ in }\Oref.\label{eq:weak_strong_divfree}
\end{align}

The viscous, mass and pressure/incompressibility constraint bilinear forms for $H(\text{div})$-conforming HDG method following \cite{Lehrenfeld10,LS15} are given by
\begin{align}
& A_h^f(v,\Psi^v)=\sum_{T\in\T_h}\langle2\nu\varepsilon(v_T),\nabla\Psi^v_T\rangle_T-\langle 2\nu\varepsilon(v_T)n,\llbracket \Psi^{v,\tau}\rrbracket\rangle_{\partial T}-\langle 2\nu\varepsilon(\Psi^v_T)n,\llbracket v^{\tau}\rrbracket\rangle_{\partial T}-\langle\frac{\nu\alpha k^2}{h}\llbracket v^{\tau}\rrbracket,\llbracket\Psi^{v,\tau}\rrbracket\rangle_{\partial T},\nonumber\\
& M^f_h(v,\Psi^v)=\langle v_T,\Psi^v_T\rangle_{\Omega},\nonumber\\
& D^f_h(v,p)=-\langle p,\Div{v_T}\rangle_{\Omega},
\end{align}
where $\llbracket v^{\tau}\rrbracket:=v_T^{\tau}-v_F^{\tau}$ denotes the tangential jump over the interfaces, $u^{\tau}:=u-(u\cdot n)n$ the tangential component, and $k$ the used polynomial order for the velocity. Note that the stability parameter $\alpha$ has to be chosen sufficiently large to obtain a coercive bilinear form. The correct facet mesh-size $h$ for (an-)isotropic elements is given by the ratio of the element volume and the boundary area, $h=\frac{J}{\Jbnd}$.

For the nonlinear convection form an up-winding technique is used, where the facet variable is glued to the up-wind triangle
\begin{align}
C^f_h(v,\Psi)=& \sum_{T\in\T_h}-\langle\nabla\Psi_Tv_T,v_T\rangle_T+\langle v_T\cdot n\,v^{up},\Psi_T\rangle_{\partial T}+\langle v_T\cdot n\,(v_F-v_T)^{\tau},\Psi_F\rangle_{\partial T_{out}},
\end{align}
with the upwind variable $v^{up}$ defined as
\begin{align}
 v^{up}:= (v_T\cdot n)n+\begin{cases}
v_T^{\tau} & \text{ if }v\cdot n \geq 0\\
v_F & \text{ if } v\cdot n < 0
\end{cases}.
\end{align}
With the Stokes bilinear form defined by
\begin{align}
B_h^f(v,p,\Psi^v,\Psi^p):= A_h^f(v,\Psi^v)+D_h^f(\Psi^v,p)+D_h^f(v,\Psi^p),
\end{align}
the variational problem for the Navier--Stokes equations reads: Find $(v,p)\in V_h\times Q_h$ such that for all $t\in [0,T]$
\begin{align}
& M_h^f(\ddt{v},\Psi^v)+ B_h^f(v,p,\Psi^v,\Psi^p)+C^f_h(v,\Psi^v) = (f,\Psi^v)\qquad \forall\, (\Psi^v,\Psi^p)\in V_h\times Q_h.
\end{align}

\subsubsection{$H(\text{div})$-conforming HDG with ALE}
After the $H(\text{div})$-conforming HDG method has been introduced, the equation has to be rewritten in the ALE context. For simplification reasons we will consider only the case of two dimensions, $d=2$. We denote a variable $u$ on the reference configuration $\Oref$ by $\uref$ and on the deformed configuration $\Ophys$ by $\uphys$. The deformation function $\varphi$ can be split into the identity function, $\idop:\R^2\rightarrow\R^2$, and the displacement $\uref:\Oref\rightarrow\R^2$, $\varphi=\idop+\uref$. Let $T$ be an element of the triangulation $\T_h$ and let $\nref$ and $\tref$ denote the corresponding outer normal and tangential vector on the boundary $\partial T$, respectively. Furthermore, the boundary determinant $\Jbnd$ is given by $\Jbnd = \|\Fmat\tref\|$ on each edge. Whereas the $H(\text{div})$-conforming elements are transformed with the Piola transformation, the facet variables get transformed with the so-called covariant transformation
\begin{align}
{\vphys}_F\circ\Phi:=\Fmat^{-T}\vref_F.
\end{align}
With the transformation rules  \eqref{eq:trafo_grad_hdiv} and \eqref{eq:trafo_time_hdiv} and the following identities for the unit normal vector, normalized tangential vector, and the mesh size of element boundaries
\begin{align}
\nphys\circ\varphi =\frac{1}{\|\Fmat^{-T}\nref\|}\Fmat^{-T}\nref,&& \tphys\circ\varphi = \frac{1}{\Jbnd}\Fmat\tref,&& \Hphys = \frac{J}{\Jbnd}\Href
\end{align}
the viscous $H(\text{div})$-conforming HDG part reads
\begin{align}
A_h^f(\vref,\uref,\Psiref)&:=\sum_{\Tref\in\T_h}\langle2\nu\,\text{sym}(\nabla P[\vref_T]\Fmat^{-1}),\nabla P[\Psiref_T]\Fmat^{-1}\rangle_{\Tref}- \langle \frac{\nu \Jbnd}{\|\Fmat^{-T}\nref\|}\text{sym}(\nabla P[\vref_T]\Fmat^{-1})\Fmat^{-T}\nref,\llbracket \Psiref^{\tau}\rrbracket\rangle_{\partial \Tref}\nonumber \\
& -\langle \frac{\nu \Jbnd}{\|\Fmat^{-T}\nref\|}\text{sym}(\nabla P[\Psiref_T]\Fmat^{-1})\Fmat^{-T}\nref,\llbracket \vref^{\tau}\rrbracket\rangle_{\partial \Tref}-\langle\frac{\nu\alpha k^2\Jbnd^2}{J\Href}\llbracket \vref^{\tau}\rrbracket,\llbracket\Psiref^{\tau}\rrbracket\rangle_{\partial \Tref},\label{eq:ale_hdg_laplace}
\end{align}
where $\llbracket \vref^{\tau}\rrbracket:=(P[\vref_T]-\Fmat^{-T}\vref_F)^{\tphys}$.

In \eqref{eq:ale_hdg_laplace} we used the volume information for the Piola transformation and the normal vector transformation as we iterate over the elements $\Tref$ and thus, have access to the element information. If we would like to use strictly the boundary integrals over the edges without additional information, the deformation gradient $\Fmat$ does not have full rank anymore. Then, the normal vector and Piola transformation would read
\begin{align}
\nphys\circ\varphi=\frac{1}{\|\cof{\Fmat}\nref\|}\cof{\Fmat}\nref, \, P[u]=\frac{1}{\Jbnd}(u\cdot\nref)\nphys\circ\varphi,
\end{align}
where $\cof{\Fmat}$ denotes the cofactor matrix of $\Fmat$. Note that the transformation for the tangent vector $\tphys$ remains the same.

The ALE-transformed mass bilinear form, together with \eqref{eq:trafo_time_hdiv}, and the pressure/incompressibility constraint is given by
\begin{align}
&\bar{M}_h^f(\vref,\uref,\Psiref):=\langle JP[\vref_T],P[\Psiref_T]\rangle_{\Oref}+\sum_{\Tref\in\T_h}\langle J(\gradref\dot{\uref}\Fmat^{-1}-\tr{\gradref\dot{\uref}\Fmat^{-1}}I)P[\vref_T],P[\Psiref_T]\rangle_{\Tref},\label{eq:mass_ale}\\
&D_h^f(\vref,\pref):=-(\Div{\vref_T},\pref)_{\Oref}.\label{eq:pres_ale}
\end{align}

Note that due to \eqref{eq:piola_trafo_div} the determinants of the deformation gradient $J$ appearing in \eqref{eq:pres_ale} cancel out. Thanks to the Piola transformation property \eqref{eq:piola_trafo_div} and the exact incompressibility \eqref{eq:weak_strong_divfree} the velocity solution is guaranteed to be exact divergence free on the reference and deformed configuration.

For the convection term the mesh velocity $\dot{\varphi}=\dot{\uref}$ has to be considered in the up-wind scheme, where the difference $P[\vref]-\dot{\uref}$ is now the corresponding wind. As $\dot{\uref}\in U_h$, it is not exactly divergence free and thus, we have to add one additional term due to integration by parts from the classical to the above used convection formulation
\begin{align}
\label{eq:conv_bfi_ale}
\bar{C}^f_h(\vref,\uref,\Psiref)&:=\sum_{\Tref\in\T_h}-\langle J\nabla\Psiref_T\Fmat^{-1}(P[\vref_T]-\dot{\uref}),P[\vref_T]\rangle_{\Tref} + \langle J\tr{\gradref\dot{\uref}\Fmat^{-1}}P[\vref_T],P[\Psiref_T]\rangle_{\Tref}\nonumber\\
&+\langle \frac{\Jbnd}{\|\Fmat^{-T}\nref\|}P[\vref_T]_{ \nref}\,\vref^{up},P[\Psiref_T]\rangle_{\partial \Tref}+\langle \frac{\Jbnd}{\|\Fmat^{-T}\nref\|}P[\vref_T]_{ \nref}\,(\Fmat^{-T}\vref_F-P[\vref_T])^{\tphys},\Fmat^{-T}\Psiref_F\rangle_{\partial \Tref_{out}},
\end{align}
where the up-wind variable now reads
\begin{align}
\vref^{up}:=(\vref_T\cdot \nphys)\nphys+\begin{cases}
\vref_T^{\tphys} & \text{ if } (\vref_T-\dot{\uref})\cdot \nphys \geq 0\\
\vref_F^{\tphys} & \text{ if } (\vref_T-\dot{\uref})\cdot \nphys  < 0
\end{cases}.
\end{align}
In \eqref{eq:conv_bfi_ale} the same additional term as in \eqref{eq:mass_ale}, namely $\langle J\tr{\gradref\dot{\uref}\Fmat^{-1}}P[\vref_T],P[\Psiref_T]\rangle_{\Tref}$, appears, but with different signs. Thus, these terms cancel out and we obtain the following weak form for the $H(\text{div})$-conforming HDG-ALE Navier--Stokes equations
\begin{align}
\label{eq:ale_navstok_disc}
M_h^f(\ddt{\vref}^f,\uref^f,\Psiref^v)+B_h^f(\vref^f,\pref^f,\uref^f,\Psiref^v,\Psiref^p)+C_h^f(\vref^f,\uref^f,\Psiref^v)+\langle J\hat{\stress}^f\Fmat^{-T}\nref^f,\Psiref^v\rangle_{\Gref}=0.
\end{align}

The boundary integral of the reference fluid stress tensor $\hat{\stress}$, which arises due to integration by parts in the viscous term, over the interface $\Gref$ is transformed with Nanson's formulae
\begin{align}
\int_{\partial\Ophys}\stress\nphys\,d\sphys=\int_{\partial\Oref}J\hat{\stress}\Fmat^{-T}\nref\,d\sref
\end{align} 
and is used as preparation for the coupling part.

\subsection{Solid discretization}
For the spatial discretization of the elastic wave equation, standard $\Hone$-conforming elements for the displacement and the velocity are used:\newline
Find $(\uref^s,\vref^s)\in U_h\times U_h$ such that for all $(\Psiref^u,\Psiref^v)\in U_h\times U_h$
\begin{align}
\label{eq:sol_disc}
&\langle\ddt{\uref^s},\Psiref^u\rangle_{\Oref}=\langle\vref^s,\Psiref^u\rangle_{\Oref},\nonumber\\
&\langle\rho^s\ddt{\vref^s},\Psiref^v\rangle_{\Oref}=-\langle\Fmat\Stress^s,\nabla\Psiref^v\rangle_{\Oref}+\langle \Pmat^s\nref^s,\Psiref^v\rangle_{\Gref}.
\end{align}
We define the following forms for a more compact notation
\begin{align}
\label{eq:op_solid_disc}
& M_h^s(\uref^s,\vref^s,\Psiref^u,\Psiref^v):=\langle\rho^s\ddt{\vref^s},\Psiref^v\rangle_{\Oref}+\langle\ddt{\uref^s}-\vref^s,\Psiref^u\rangle_{\Oref},\nonumber\\
& K_h^s(\uref^s,\Psiref^u):= \langle\Fmat\Stress^s,\nabla\Psiref^u\rangle_{\Oref}.
\end{align}

\subsection{Deformation extension for mesh movement}
For the mesh velocity $\dot{\uref}$, and thus for the ALE description, a displacement field $\uref^f$ is needed on the reference fluid domain $\Oref^f$. It is artificial and has to ensure that the displacement on the interface from the solid to the fluid is continuous
\begin{align}
\uref^f = \uref^s \text{ on } \Gref.
\end{align}
Hence, the displacement $\uref^f$ on the fluid domain can be seen as an extension of $\uref^s$, which is realized by an auxiliary mapping $\mathcal{A}:\Gref\rightarrow\Oref^f$.

There is a amount of possibilities to choose the deformation extension problem. E.g., in \cite{HEL03} the biharmonic problem was investigated and in \cite{Wick011} linear extensions were compared. In this paper, however, a nonlinear elasticity problem with a Neo--Hookean like material law \cite{Br2013} is considered: Find $\uref^f\in U_h\cap\Oref^f$ with $\uref^f=\uref^s$ on $\Gref$, such that for all $\Psiref\in U_h\cap\Oref^f$
\begin{align}
\langle\beta\mu(I-\det(\Cmat)^{-\frac{\lambda}{2\mu}}\Cmat^{-1}),\Fmat^T\gradref\Psiref\rangle_{\Oref^f} = 0,
\end{align}
and define
\begin{align}
\label{eq:def_ext_bfi}
N_h^f(u,\Psi):=\langle\beta\mu(I-\det(\Cmat)^{-\frac{\lambda}{2\mu}}\Cmat^{-1}),\Fmat^T\gradref\Psi\rangle_{\Oref^f}.
\end{align}
Here, $\beta:\Oref^f\rightarrow\R$ denotes a spatial dependent coefficient function given by
\begin{align}
\label{eq:beta_def_ext}
\beta(\xref) = \frac{c}{\sqrt{|\text{dist}(\xref,\Gref)|^2+\varepsilon}},\,\, \text{dist}(\xref,\Gref)=\min\limits_{a\in\Gref}\|\xref-a\|,
\end{align} 
with $1 \gg c > 0$ and a small regularization parameter $\varepsilon$. It stiffens the extension problem near the interface, where the deformations are expected to be critical.

The artificial material parameters $\mu$ and $\lambda$ in \eqref{eq:def_ext_bfi} can be adopted. In numerical experiments, however, we observed that setting both to $1$ is already satisfying. In a monolithic approach the deformation extension also infects the solid equation, as it can be interpreted as a boundary condition. To minimize this unintended effect, the parameter $c$ in \eqref{eq:beta_def_ext} has to be chosen sufficiently small.

\subsection{Coupling}
\label{subsec:coupling}
To couple both equations, the displacement and velocity of the fluid and solid have to be continuous over the interface $\Gref$. As the displacement is discretized globally by $\Hone$-conforming elements, the continuity follows immediately. But the fluid and solid velocity live in two different finite element spaces
\begin{align}
\vref^f\in V_h=W_h\times F_h\quad\text{ and }\quad \vref^s\in U_h.
\end{align}
To enforce continuity over the interface we use Lagrange multipliers, which live in the $\Ltwo[\Gref]$-conforming space on the interface
\begin{align}
\mathcal{L}_h:= [\Pi^k(\hat{\Gamma})]^d
\end{align}
and the corresponding equations in the deformed configuration read
\begin{align}
&\langle ({\vphys}_T^f-\vphys^s)^{\nphys},{\Psiphys}^{\lambda_1}\rangle_{\Gphys}+\langle ({\Psiphys}_T^{v,f}-\Psiphys^{v,s})^{\nphys},{\lphys}_1\rangle_{\Gphys}=0,\nonumber\\
&\langle({\vphys}_F^f-\vphys^s)^{\tphys},{\Psiphys}^{\lambda_2}\rangle_{\Gphys} +\langle ({\Psiphys}_F^{v,f}-\Psiphys^{v,s})^{\tphys},{\lphys}_2\rangle_{\Gphys} =0.
\end{align}
In the ALE context these equations transform to
\begin{align}
\label{eq:ale_interface}
&\langle \Jbnd(P[\vref_T^f]-\vref^s)^{\nphys},\Psiref^{\lambda_1}\rangle_{\Gref}+\langle \Jbnd(P[\Psiref_T^{v,f}]-\Psiref^{v,s})^{\nphys},\lref_1\rangle_{\Gref}=0,\nonumber\\
&\langle \Jbnd(\Fmat^{-T}\vref_F^f-\vref^s)^{\tphys},\Psiref^{\lambda_2}\rangle_{\Gref}+\langle \Jbnd(\Fmat^{-T}\Psiref_F^{v,f}-\Psiref^{v,s})^{\tphys},\lref_2\rangle_{\Gref}=0.
\end{align}
We define the bilinear form
\begin{align}
\label{eq:lag_mult}
 L_h(\vref^f,\vref^s,\lref,\Psiref^f,\Psiref^s,\Psiref^{\lambda}):=L^{n}_h(\vref^f,\vref^s,\lref,\Psiref^f,\Psiref^s,\Psiref^{\lambda})+L^{\tau}_h(\vref^f,\vref^s,\lref,\Psiref^f,\Psiref^s,\Psiref^{\lambda}),
\end{align}
with
\begin{align}
L^n_h(\vref^f,\vref^s,\lref,\Psiref^f,\Psiref^s,\Psiref^{\lambda}):=\langle \Jbnd(P[\vref_T^f]-\vref^s)^{\nphys},\Psiref^{\lambda_1}\rangle_{\Gref}+\langle \Jbnd(P[\Psiref_T^{v,f}]-\Psiref^{v,s})^{\nphys},\lref_1\rangle_{\Gref}
\end{align}
and analogously $L^{\tau}_h$.

Due to the continuity conditions for the displacement and the velocity we can define the following global functions $\uref$, $\vref$ by
\begin{align}
\uref(\xref):=\begin{cases}
\uref^f(\xref) & \xref\in\Oref^f\\
\uref^s(\xref) & \xref\in\Oref^s
\end{cases},\, \vref(\xref):=\begin{cases}
\vref^f(\xref) & \xref\in\Oref^f\\
\vref^s(\xref) & \xref\in\Oref^s
\end{cases}.
\end{align}

As a monolithic approach will be used, we have to sum all equations and solve them at once, i.e., with \eqref{eq:ale_navstok_disc}, \eqref{eq:sol_disc}, \eqref{eq:def_ext_bfi}, and \eqref{eq:lag_mult} the complete problem reads:\newline
Find $(\uref,\pref,\vref,\lref)\in [U_h,V_h]\times Q_h\times V_h\times \mathcal{L}_h$ such that for all $(\Psiref^u,\Psiref^p,\Psiref^v,\Psiref^{\lambda})\in [U_h,V_h]\times Q_h\times V_h\times \mathcal{L}_h$ and all $t\in [0,T]$
\begin{align}
\label{eq:all_terms_fsi_spat}
& A_h^f(\vref,\uref,\Psiref^v)+B_h^f(\vref,\pref,\uref,\Psiref^v,\Psiref^p)+C_h^f(\vref,\uref,\Psiref^v)+M_h^f(\ddt{\vref},\uref,\Psiref^v)+M_h^s(\ddt{\uref},\vref,\Psiref^u,\Psiref^v)+K_h^s(\uref,\Psiref^v)\nonumber\\
&\,+L_h(\vref^f,\vref^s,\lref,\Psiref^f,\Psiref^s,\Psiref^{\lambda})+N^f_h(\uref,\Psiref^u)+\langle \Pmat^s\nref^s,\Psiref^v\rangle_{\Gref}+\langle J\hat{\stress}^f\Fmat^{-T}\nref^f,\Psiref^v\rangle_{\Gref}=0.
\end{align}
To ensure the balance of forces on the interface we can simply neglect the two interface integrals in \eqref{eq:all_terms_fsi_spat}, called strongly coupled
approach \cite{WGR07,TSS06}. Thus, the condition is handled implicitly in a natural way.

\subsection{Time discretization}
From now on, we will neglect the sub- and superscripts, which refer to the reference or deformed quantity. For a complete discretization we are going to use the method of lines. Thus, after the spatial discretization is done, the L-stable second-order SDIRK  Runge--Kutta method \cite{alex77} is applied on scheme \eqref{eq:all_terms_fsi_spat}.

Only the pressure/incompressibility constraint, the velocity continuity condition and the deformation extension are handled completely implicit.

Other possible choices for the temporal discretization are e.g. backwards difference formulas (BDF) \cite{SSJ16} or Fractional-Step-$\Theta$ methods \cite{Turek2010}.

\section{Numerical examples}
The performance of the presented method is tested with the following two-dimensional benchmarks purposed by Turek and Hron \cite{TH06,Turek2010}, which are based on the configurations of the classical flow around cylinder CFD benchmark in \cite{ST96}.

\subsection{Implementation aspects}
For all numerical experiments the open source finite element library NETGEN \cite{Sch97} and NGSolve\footnote{www.ngsolve.org} \cite{Sch14} is used. The nonlinear problem is solved by Newton's method and the therein arising non-symmetric linearized problems with the direct solver UMFPACK\footnote{http://faculty.cse.tamu.edu/davis/suitesparse.html} \cite{DD97}.

Computing the directional derivatives of \eqref{eq:all_terms_fsi_spat} is quite involved due to the nonlinearties arising from the ALE transformations. NGSolve supports symbolic integrators with automatic exact differentiation such that one can use \eqref{eq:all_terms_fsi_spat} directly - there is no need to compute the stiffness matrix by hand. Another approach is to use Newton's method as a fix-point iteration replacing the (nonlinear) terms from the transformations by the terms from the previous time step or the previous Newton iteration. In numerical experiments we observed that this yields the same results, with the advantage of a speed-up.

The polynomial order for the pressure is of one degree less than for the velocity \eqref{eq:pressure_space}. Due to the construction of the $H(\text{div})$-conforming finite elements in NGSolve \cite{Zaglmayr06,Lehrenfeld10,LLS18} it is possible to neglect all high-order $H(\text{div})$ velocity basis functions with non-zero divergence, as their coefficients would be zero anyway. Thus, only piece-wise constants, $k=0$, have to be used for the pressure reducing the number of degrees of freedom (dofs) \cite{LS15}. Note that this has no influence to the quality of the velocity solution and  one can recover the high-order approximation of the pressure by solving cheap element-wise problems as a post-processing step.

We apply static condensation on element level to eliminate the internal bubbles reducing the number of dofs further, which has an enormous impact for high polynomial degrees.

The stability parameter $\alpha$ in \eqref{eq:ale_hdg_laplace} is set to $5$ for all benchmarks. For the regularization parameter we use $\varepsilon=10^{-12}$ and for the constant $c$ in \eqref{eq:beta_def_ext} $c=5\times 10^{-17}$ for the stationary case and $c=2\times 10^{-14}$ for the others.

\subsection{Geometry}
\begin{figure}
	\centering
	\begin{tabular}{cc}
	\includegraphics[width=0.48\textwidth]{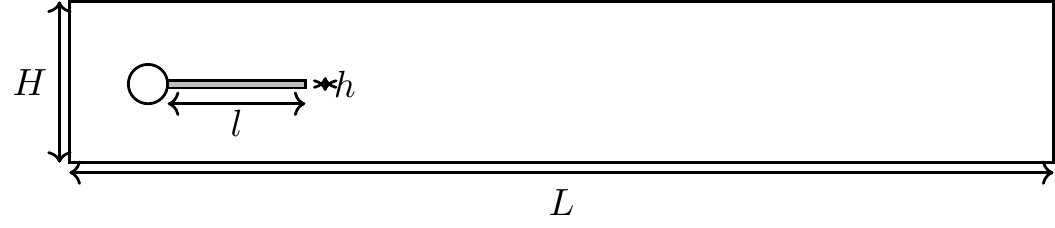}\hspace*{0.1cm}&
	\includegraphics[width=0.44\textwidth]{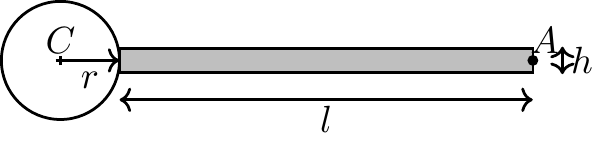}
	\end{tabular}
	
	\caption{Full channel and flag geometry of the benchmark.}
	\label{fig:geom_channel}
\end{figure}

The benchmark consists of a channel with a cylinder, placed slightly non-symmetric. For the FSI benchmarks an elastic flag is attached at the end of this cylinder. The geometry data can be found in Table \ref{tab:turek_geom_param} and seen in Figure \ref{fig:geom_channel}.

\begin{table}
\centering
\begin{tabular}{lrc}
\hline
Parameter &  & value [m] \\
\hline
channel length & $L$ & $2.5$ \\
channel width & $H$ & $0.41$ \\
cylinder position & $C$ & $(0.2,0.2)$ \\
cylinder radius & $r$ & $0.05$ \\
solid structure length & $l$ & $0.35$ \\
solid structure height & $h$ & $0.02$ \\
reference point (at $t=0$) & $A$ & $(0.6,0.2)$ \\
\hline
\end{tabular}
\caption{Geometry parameters of FSI benchmarks.}
\label{tab:turek_geom_param}
\end{table}

\subsection{Boundary data, initial condition and quantities of interest }
A parabolic inflow profile is prescribed at the left boundary of the channel by the function
\begin{align}
\label{eq:td_vel_profile}
v^f(0,y,t)=\begin{cases}
v^f(0,y)\frac{1-\cos(\frac{\pi}{2}t)}{2} & \text{if } t < 2,\\
v^f(0,y) & \text{otherwise},
\end{cases}
\end{align}
where
\begin{align}
\label{eq:vel_profile}
v^f(0,y)=6\overline{U}\frac{y(H-y)}{H^2}=\overline{U}\frac{6}{0.1681}y(0.41-y)
\end{align}
is chosen in such a way that $\overline{U}$ and $1.5\overline{U}$ are the mean and maximal velocities, respectively. For the outflow boundary we choose the do-nothing condition, $\sigma_n^f=0$, and on the other boundaries the no-slip condition, $v^f_D=0$.

One quantity of comparison is the displacement of the control point $A$ on the right end of the elastic flag. Furthermore, the drag and lift forces over the cylinder and the interface are computed by
\begin{align}
(F_D,F_L) = \int_{\Sphys}\sigma n\,d\sphys,
\end{align}
where $\Sphys$ denotes the boundary between the fluid domain and the obstacle together with the elastic flag.

For the FSI benchmarks three different settings for the parameters are used, which are listed in Table \ref{tab:param_fsi_tests}. In the case of the FSI 1 benchmark the solution converges to a steady state, whereas in the other two settings the solutions become periodically.
\begin{table}
	\centering
	\begin{tabular}{lrrr}
		\hline
		Parameter & $\quad$FSI 1 & $\quad$FSI 2 & $\quad$FSI 3 \\
		\hline
		$\rho^s\,[10^3]$ & $1$ & $10$ & $1$ \\
		$\nu^s$ & $0.4$ & $0.4$ & $0.4$ \\
		$\mu^s\,[10^6]$ & $0.5$ & $0.5$ & $2$ \\
		\hline
		$\rho^f\,[10^3]$ & $1$ & $1$ & $1$ \\
		$\nu^f\,[10^{-3}]$ & $1$ & $1$ & $1$ \\
		\hline
		$\overline{U}$ & $0.2$ & $1$ & $2$ \\
		\hline
	\end{tabular}
	\caption{Parameters for the FSI benchmark tests.}
	\label{tab:param_fsi_tests}
\end{table}

\subsection{Mesh}
\begin{figure}
	\centering
	\includegraphics[width=0.49\textwidth]{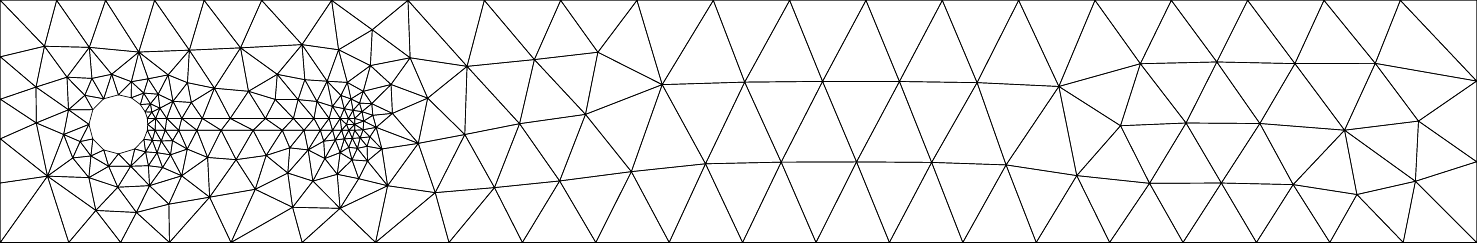}
	\includegraphics[width=0.49\textwidth]{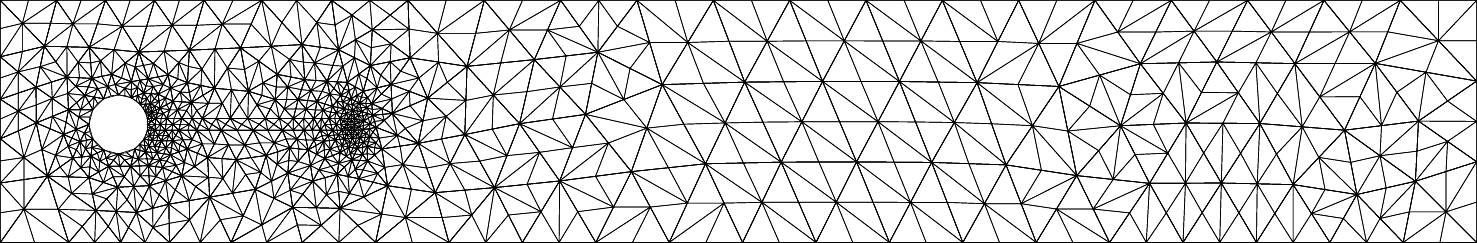}
	\includegraphics[width=0.49\textwidth]{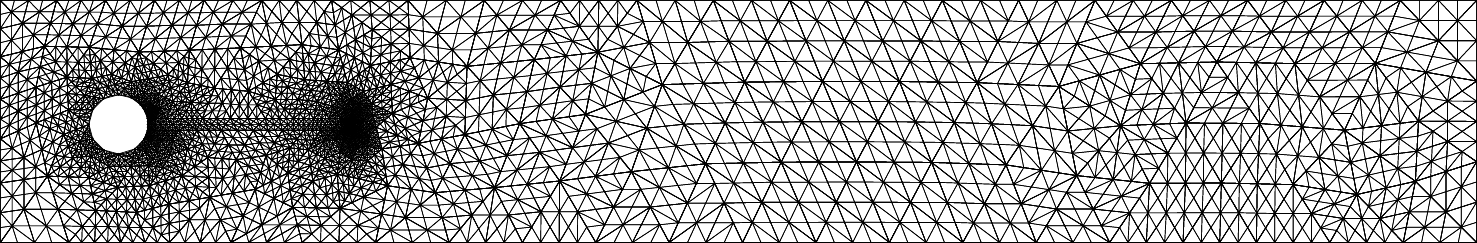}
	\caption{Coarse mesh and first two uniform refinement levels.}
	\label{fig:mesh_coarse_uni_ref}
\end{figure}

The mesh is generated automatically from the geometry by NETGEN. The coarsest level and the first two uniform refinement levels are depicted in Figure \ref{fig:mesh_coarse_uni_ref}. Due to the cylindrical obstacle we use curved boundary elements of the same order as the velocity and the displacement. The grid at the right part of the channel is slightly coarser than the important areas around the elastic flag and obstacle.

\begin{figure}
	\centering
	\includegraphics[width=0.49\textwidth]{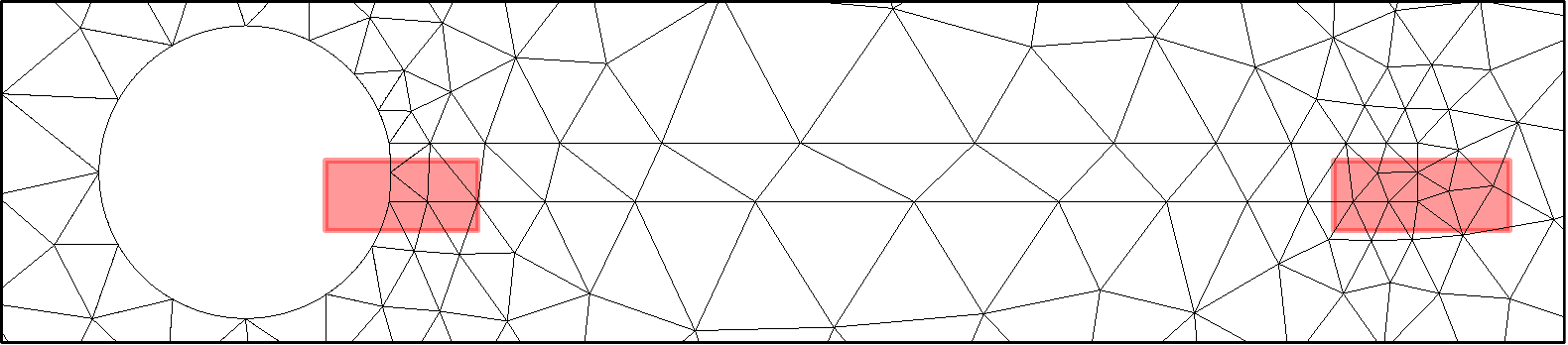}
	
	\includegraphics[width=0.14\textwidth]{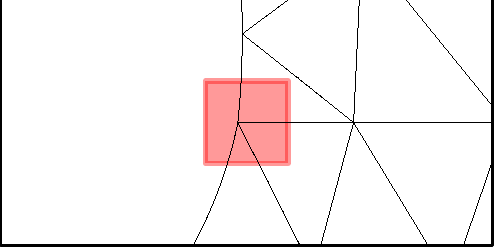}
	\includegraphics[width=0.07\textwidth]{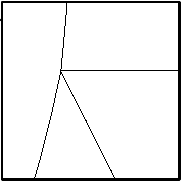}
	\hspace*{0.675cm}
	\includegraphics[width=0.14\textwidth]{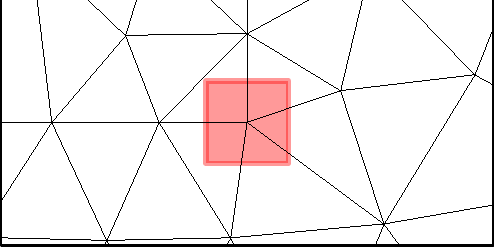}
	\includegraphics[width=0.07\textwidth]{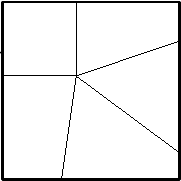}
	\caption{Zooming towards to two of the four singularities in the coarsest mesh.}
	\label{fig:mesh_zoom_sing}
\end{figure}

\begin{figure}
	\centering
\begin{tabular}{ccc}
	\begin{minipage}[b]{.43\textwidth}
		\includegraphics[width=1\textwidth]{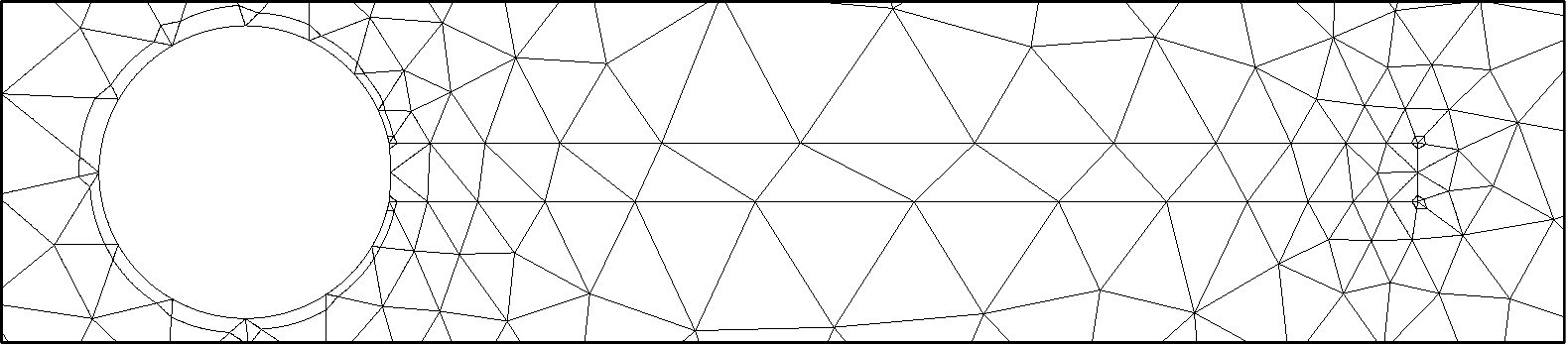}
		\includegraphics[width=1\textwidth]{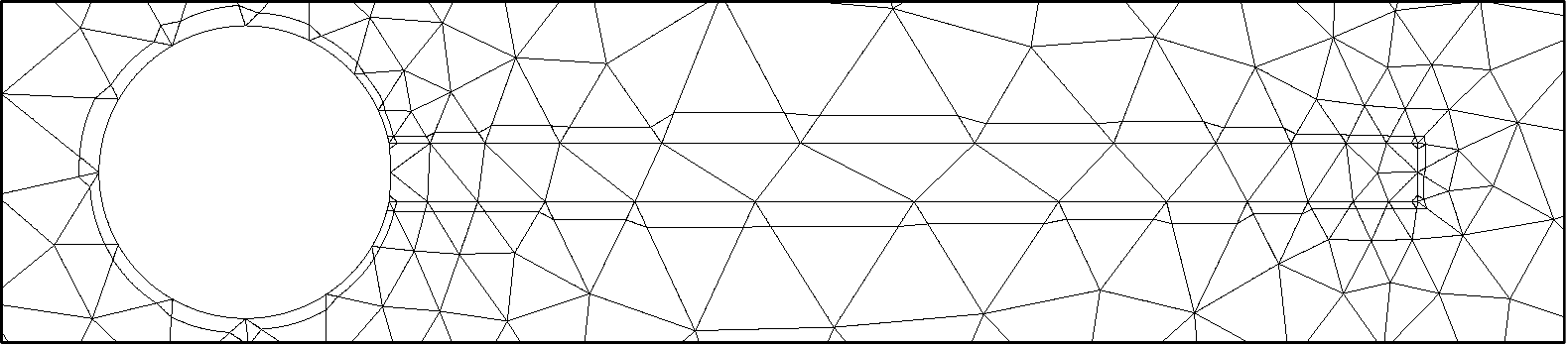}
	\end{minipage}
	\includegraphics[width=.19\textwidth]{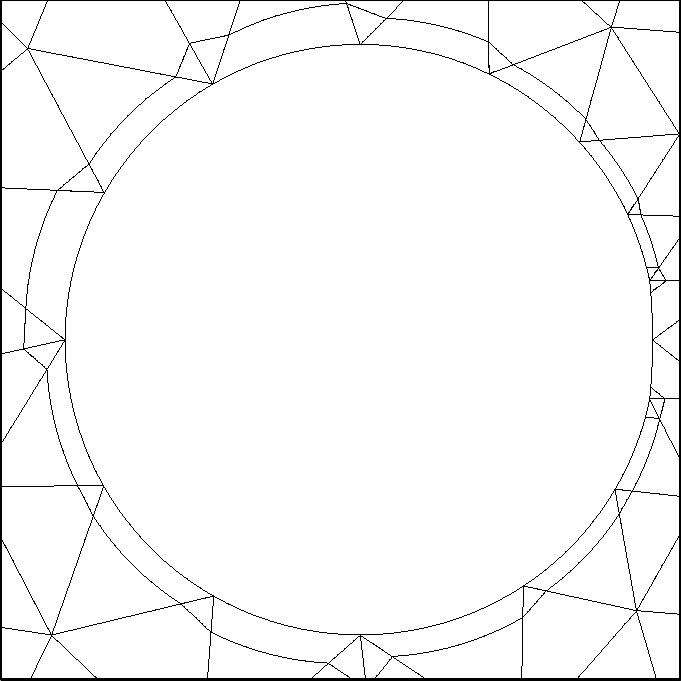}&
		\begin{minipage}[b]{.36\textwidth}
			\includegraphics[width=1\textwidth]{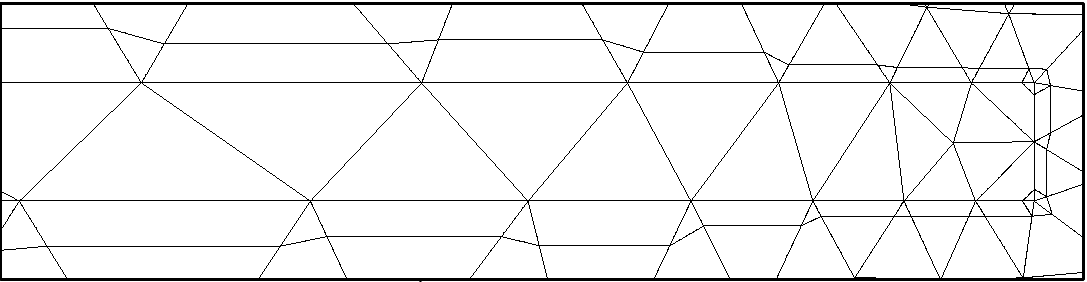}\hfill
			\includegraphics[width=1\textwidth]{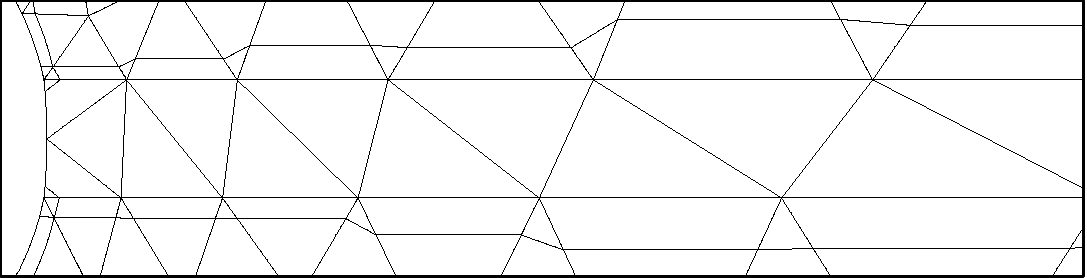}
		\end{minipage}
\end{tabular}
	\caption{Hp-refinement with boundary layers at cylinder and with boundary layers at cylinder and flag (left), zooming to cylinder (middle), and zooming to flag (right).}
	\label{fig:mesh_zoom_bl}
\end{figure}

In the numerical experiments we observed that using polynomial degree $k$ does not lead to the optimal convergence rate $\mathcal{O}(h^k)$, see results in Tables \ref{tab:result_uni_p2_fsi1} -- \ref{tab:result_uni_p5_fsi1}. Here, $h$ denotes the mesh-size of the quasi-uniform triangulations. With a Zienkiewicz--Zhu (ZZ) a-posterior error estimator \cite{zienkiewicz87,wu90} we could identify four singularities: the two right corners of the flag due to the non-convexity of the fluid domain and the corners on the left, where the flag is fixed, see Figure \ref{fig:mesh_zoom_sing}. Furthermore, in the FSI 2  benchmark a boundary layer around the obstacle needs to be resolved and for the FSI 3 benchmark additionally the boundary layer around the flag. Thus, to repair the convergence rates, we use an hp-refinement strategy with a geometric refinement-factor of $0.2$, where we refine around the singularities and twice at the boundary layers, see Figure \ref{fig:mesh_zoom_bl}, and increase the polynomial degree away from the singularities. It is also possible to use uniform polynomial degree to obtain the same asymptotic rate, however, this would lead to more degrees of freedom than needed.

\begin{figure}
	\centering
	\includegraphics[width=0.12\textwidth]{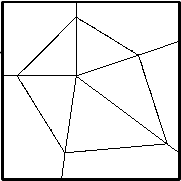}
	\includegraphics[width=0.12\textwidth]{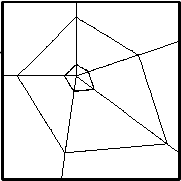}
	\includegraphics[width=0.12\textwidth]{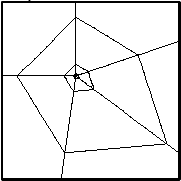}
	\includegraphics[width=0.12\textwidth]{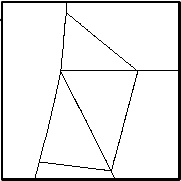}
	\includegraphics[width=0.12\textwidth]{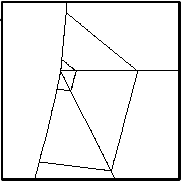}
	\caption{Refinement at singularities. From left to right: first three geometric refinement steps at the right singularity and first two refinements at the left one.}
	\label{fig:mesh_zoom_sing_ref}
\end{figure}

As the singularities on the left side have less impact to the solution, we refine them only at refinement step three, five and eight for the FSI 1 benchmark and the first two times for the instationary examples, whereas the right corners get refined at every step, see Figure \ref{fig:mesh_zoom_sing_ref}.

Due to the pressure robustness of the $H(\text{div})$-conforming method, we can neglect the pressure $p$ and only use the gradient of the velocity $u^f$ to estimate the fluid error \cite{LS15}. For the solid and deformation extension error we interpolate the first Piola--Kirchhoff stress tensor.

\begin{table}
\centering
\begin{tabular}{c|ccccccccc}
lvl & 0 & 1 & 2 & 3 & 4 & 5 & 6 & 7 & 8 \\
\hline
\hline
uni & 397 & 1588 & 6352 & 25408&&&&&\\
\hline
bl1 & 397 & 441 & 487 & 497 & & & &&\\
\hline
bl2 & 397 & 492 & 595 & 605 & & & &&\\
\hline
hp & 397 & 407 & 417 & 433 & 443 & 459 & 469 & 479 & 495
\end{tabular}
\caption{Refinement levels and number of elements for uniform refinement, hp refinement with boundary layers around circle, hp refinement with boundary layers around cylinder and flag (each 4 levels), and hp refinement without boundary layers (8 levels).}
\label{tab:level_uni_ref}
\end{table}

\begin{table}
\centering
\begin{tabular}{c|c|c|c|c|c}
lvl & ndof & $u_x [\times 10^{-5}]$ & $u_y [\times 10^{-4}]$ & drag & lift \\
\hline
0 & 7257 & 2.2390 & 8.7913 & 14.3194 & 0.79752 \\
1 & 28242 & 2.2721 & 8.0243 & 14.2829 & 0.75635 \\
2 & 110010 & 2.2692 & 8.2039 & 14.2934 & 0.76405 \\
3 & 432738 & 2.2694 & 8.2041 & 14.2939 & 0.76442 \\
\end{tabular}
\caption{Results for FSI 1 benchmark with uniform refinement p2 HDG.}
\label{tab:result_uni_p2_fsi1}
\end{table}

\begin{table}
\centering
\begin{tabular}{c|c|c|c|c|c}
lvl & ndof & $u_x [\times 10^{-5}]$ & $u_y [\times 10^{-4}]$ & drag & lift \\
\hline
0 & 11595 & 2.2755 & 8.1076 & 14.2936 & 0.77417 \\
1 & 44858 & 2.2702 & 8.1965 & 14.2941 & 0.76458 \\
2 & 175002 & 2.2695 & 8.2065 & 14.2942 & 0.76446 \\
3 & 689762 & 2.2696 & 8.2009 & 14.2941 & 0.76439 \\
\end{tabular}
\caption{Results for FSI 1 benchmark with uniform refinement p3 HDG.}
\label{tab:result_uni_p3_fsi1}
\end{table}
\begin{table}
	\centering
	\begin{tabular}{c|c|c|c|c|c}
lvl & ndof & $u_x [\times 10^{-5}]$ & $u_y [\times 10^{-4}]$ & drag & lift \\
\hline
0 & 17153 & 2.2716 & 8.2184 & 14.2958 & 0.76482 \\
1 & 66354 & 2.2696 & 8.2092 & 14.2944 & 0.76447 \\
2 & 259514 & 2.2697 & 8.2023 & 14.2942 & 0.76440 \\
3 & 1024866 & 2.2697 & 8.1988 & 14.2941 & 0.76437 \\
\end{tabular}
	\caption{Results for FSI 1 benchmark with uniform refinement p4 HDG.}
	\label{tab:result_uni_p4_fsi1}
\end{table}
\begin{table}
\centering
\begin{tabular}{c|c|c|c|c|c}
lvl & ndof & $u_x [\times 10^{-5}]$ & $u_y [\times 10^{-4}]$ & drag & lift \\
\hline
0 & 23931 & 2.2709 & 8.2135 & 14.2959 & 0.76454 \\
1 & 92730 & 2.2697 & 8.2053 & 14.2943 & 0.76443 \\
2 & 363546 & 2.2697 & 8.2003 & 14.2941 & 0.76438 \\
3 & 1438050 & 2.2697 & 8.1978 & 14.2940 & 0.76436 \\
\end{tabular}
\caption{Results for FSI 1 benchmark with uniform refinement p5 HDG.}
\label{tab:result_uni_p5_fsi1}
\end{table}

\begin{table}
\centering
\begin{tabular}{c|c|c|c|c|c}
nel & ndof & $u_x [\times 10^{-5}]$ & $u_y [\times 10^{-4}]$ & drag & lift \\
\hline
0 & 7257 & 2.2391 & 8.79420 & 14.3195 & 0.79791 \\
1 & 12021 & 2.2734 & 8.06994 & 14.2904 & 0.77411 \\
2 & 18245 & 2.2696 & 8.19171 & 14.2931 & 0.76478 \\
3 & 26651 & 2.2697 & 8.19438 & 14.2939 & 0.76434 \\
4 & 36227 & 2.2696 & 8.19507 & 14.2940 & 0.76436 \\
5 & 48393 & 2.2697 & 8.19540 & 14.2940 & 0.76434 \\
6 & 61697 & 2.2697 & 8.19540 & 14.2940 & 0.76434 \\
7 & 76861 & 2.2697 & 8.19541 & 14.2940 & 0.76434 \\
8 & 95447 & 2.2697 & 8.19543 & 14.2940 & 0.76434 \\
\end{tabular}
\caption{Results for FSI 1 benchmark with hp-refinement HDG.}
\label{tab:result_hp_fsi1}
\end{table}

\begin{figure}
\centering
\includegraphics[width=0.47\textwidth]{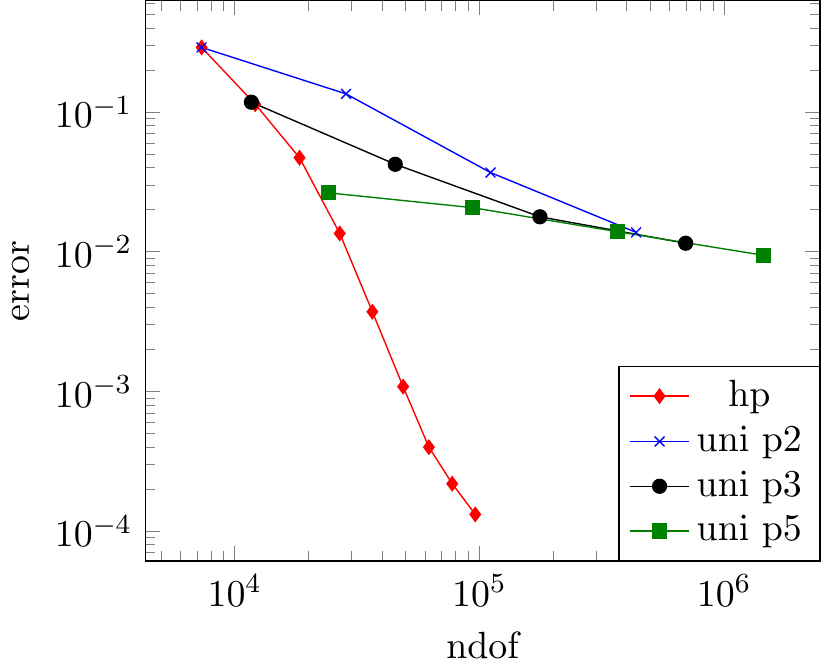}
\caption{Estimated $L^2$-error for uniform refinement with polynomial degree $2$, $3$, and $5$ and hp-refinement strategies for FSI 1 benchmark after the solution became stationary.}
\label{fig:err_uni_hp_fsi1}
\end{figure}

\begin{figure}
\centering
\includegraphics[width=0.4\textwidth]{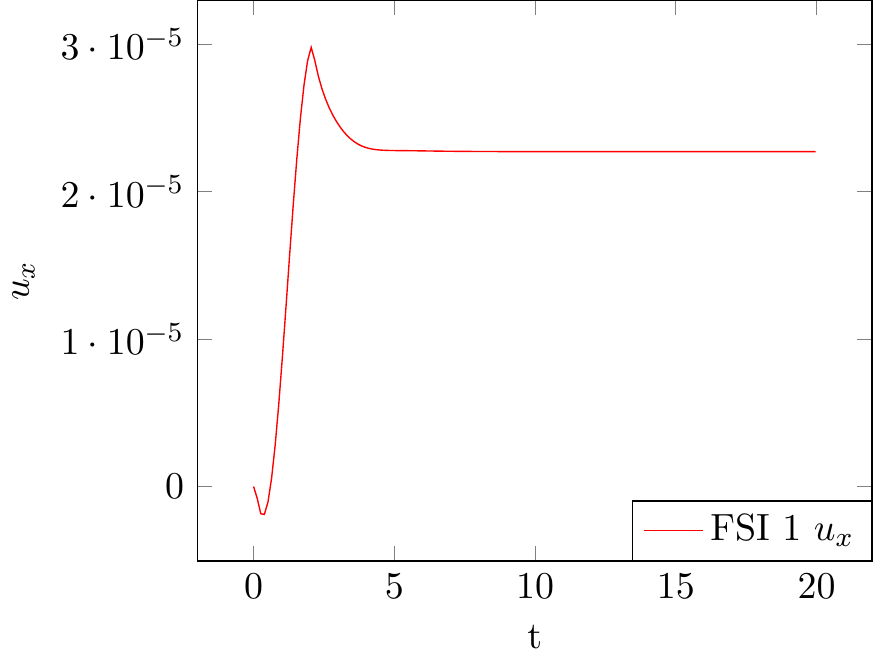}
\includegraphics[width=0.4\textwidth]{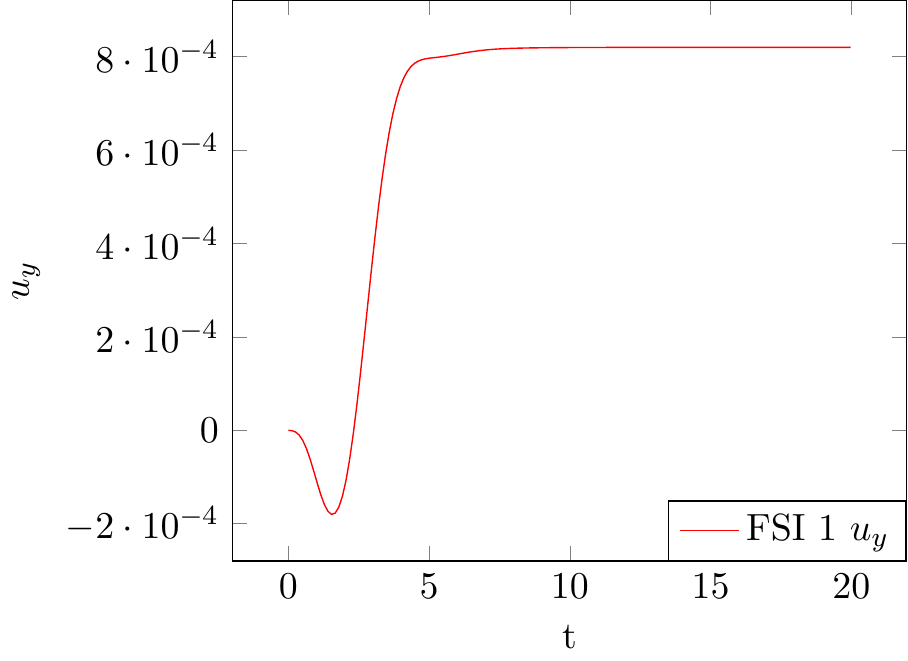}
\includegraphics[width=0.4\textwidth]{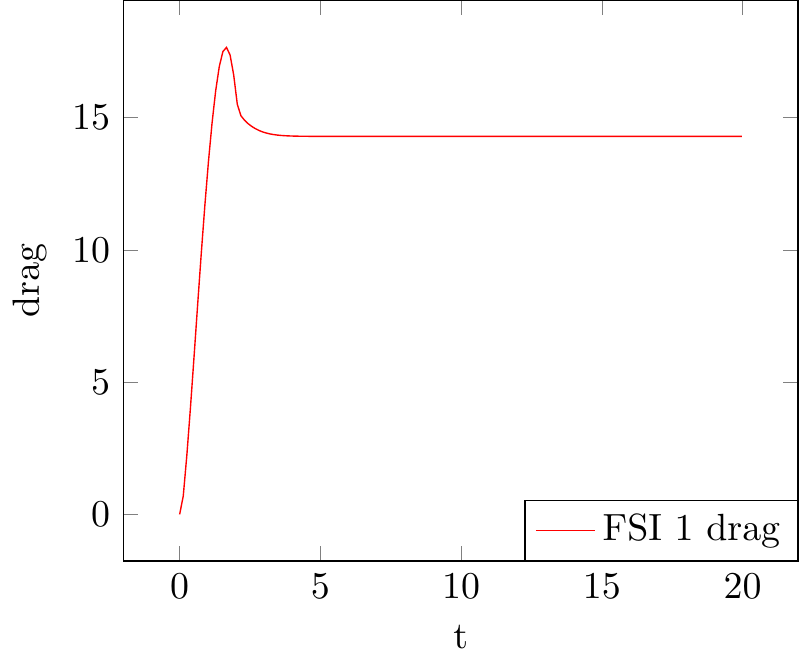}
\includegraphics[width=0.4\textwidth]{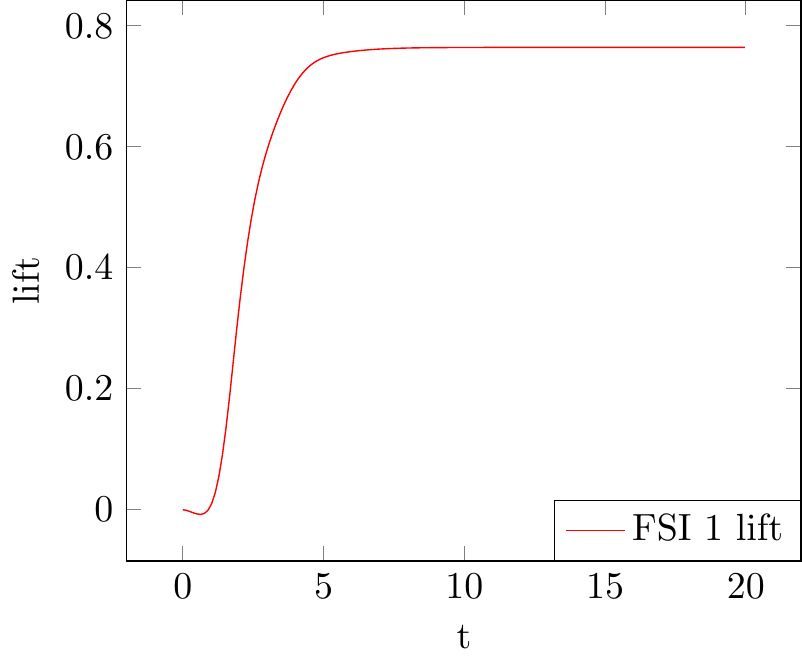}
\caption{Displacements $u_x$, $u_y$ at control point $A$ and drag, lift forces for FSI 1 benchmark.}
\label{fig:fsi1_plots}
\end{figure}

For the stationary FSI 1 benchmark we start with uniform polynomial degree two for the hp-refinement strategy. The results for hp-refinement is given in Table \ref{tab:result_hp_fsi1} and the estimated $L^2$-error of the uniform $h$ and hp-refinement strategy with respect to the number of degrees of freedom -see Table \ref{tab:level_uni_ref}- is shown in Figure \ref{fig:err_uni_hp_fsi1}. One can observe that the uniform $h$-refinement strategy leads to a loss of the optimal convergence rate when the error at the singularities becomes dominant, whereas the hp-refinement does not suffer from this behavior as the singularities are resolved. The values given in Table \ref{tab:result_hp_fsi1} agree with the comparison results in \cite{THR10}. In Figure \ref{fig:fsi1_plots} is reported how the values become stationary after $t=8$.

\begin{figure}
	\centering
	\includegraphics[width=0.80\textwidth]{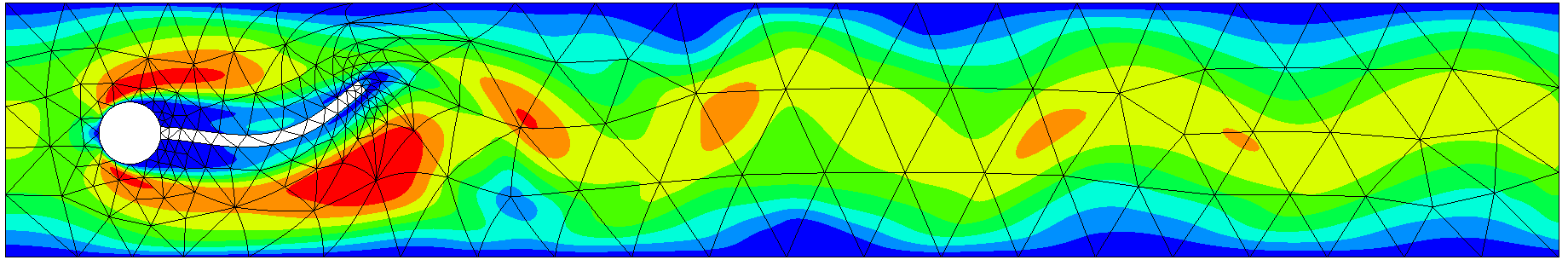}
	\caption{FSI 2 snapshot at $t=10.24$ for the coarsest mesh and $\tau=0.004$.}
	\label{fig:fsi2_large_def}
\end{figure}

\begin{figure}
\centering
\includegraphics[width=0.4\textwidth]{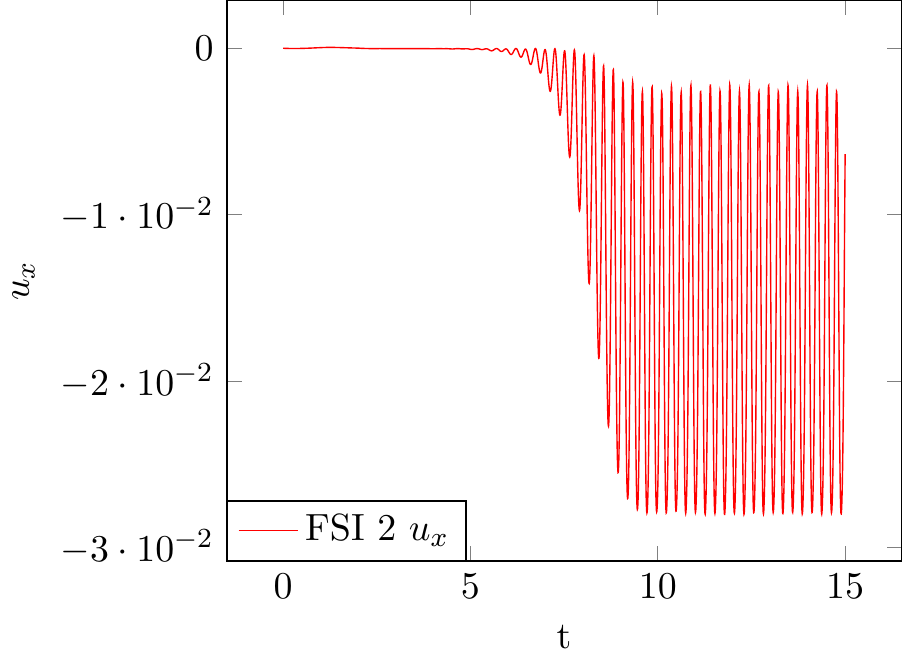}
\includegraphics[width=0.4\textwidth]{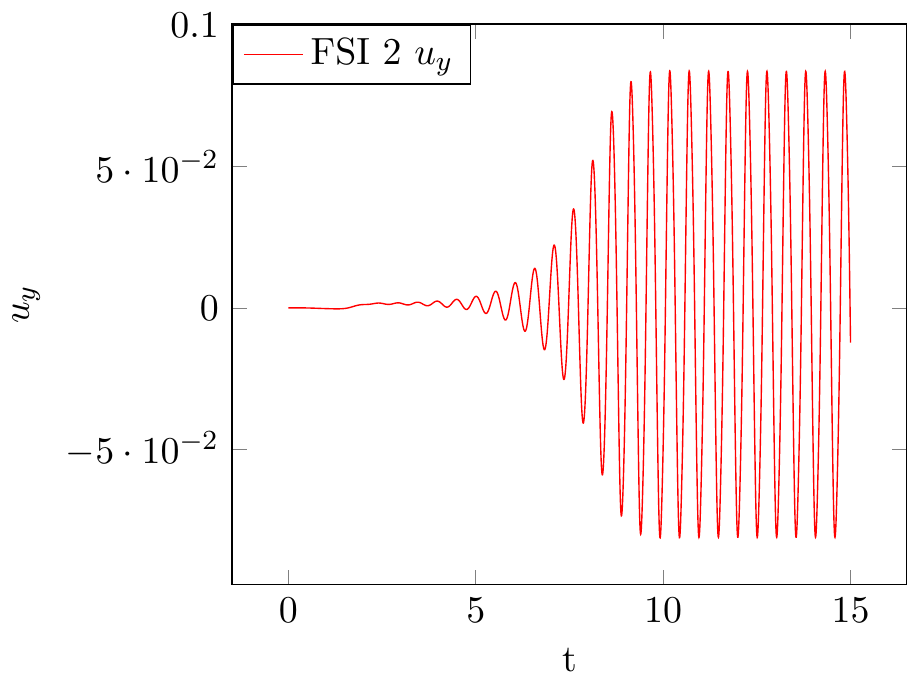}
\includegraphics[width=0.4\textwidth]{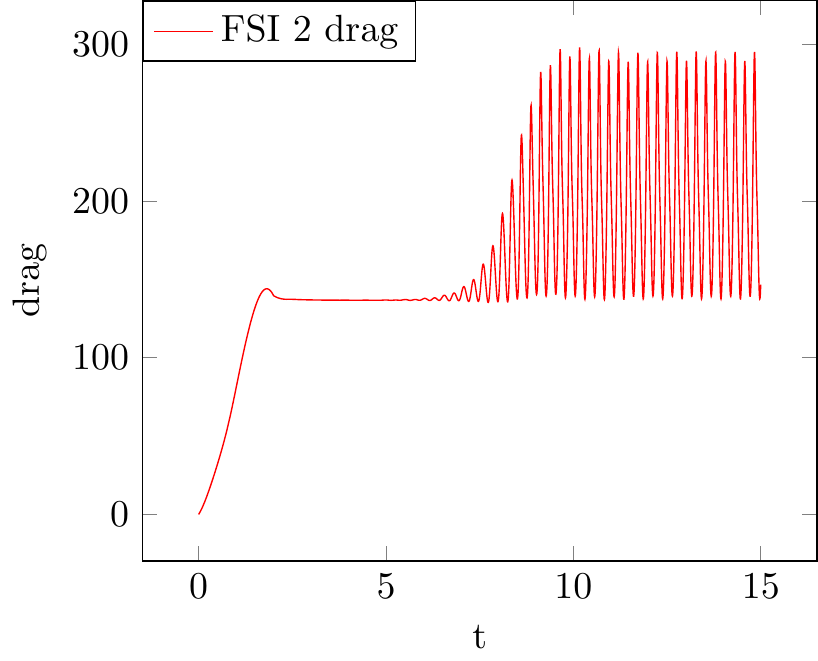}
\includegraphics[width=0.4\textwidth]{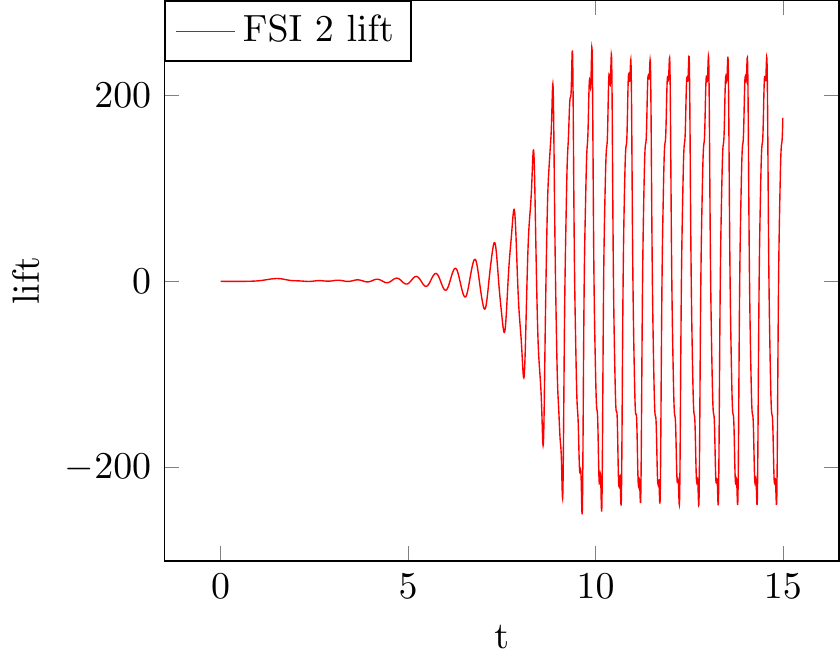}
\caption{Displacements $u_x$, $u_y$ at control point $A$ and drag, lift forces for FSI 2 benchmark.}
\label{fig:fsi2_plots}
\end{figure}

The FSI 2 benchmark involves quite large deformations of the elastic beam, see Figure \ref{fig:fsi2_large_def}. As we use three hp-refinement stages the initial polynomial degree is set to four. After time $t=12$ the solutions start to become oscillating, see Figure \ref{fig:fsi2_plots}. The results, which can be found in Table \ref{tab:result_bl_hp_hdiv_fsi2}, correspond with \cite{Turek2010,Wick13}.

\begin{table}
\centering
\input{fsi2_final_bl_minmax.out}
\caption{Results for FSI 2 with hp-refinement HDG and boundary layers around cylinder.}
\label{tab:result_bl_hp_hdiv_fsi2}
\end{table}

\begin{figure}
\centering
\includegraphics[width=0.4\textwidth]{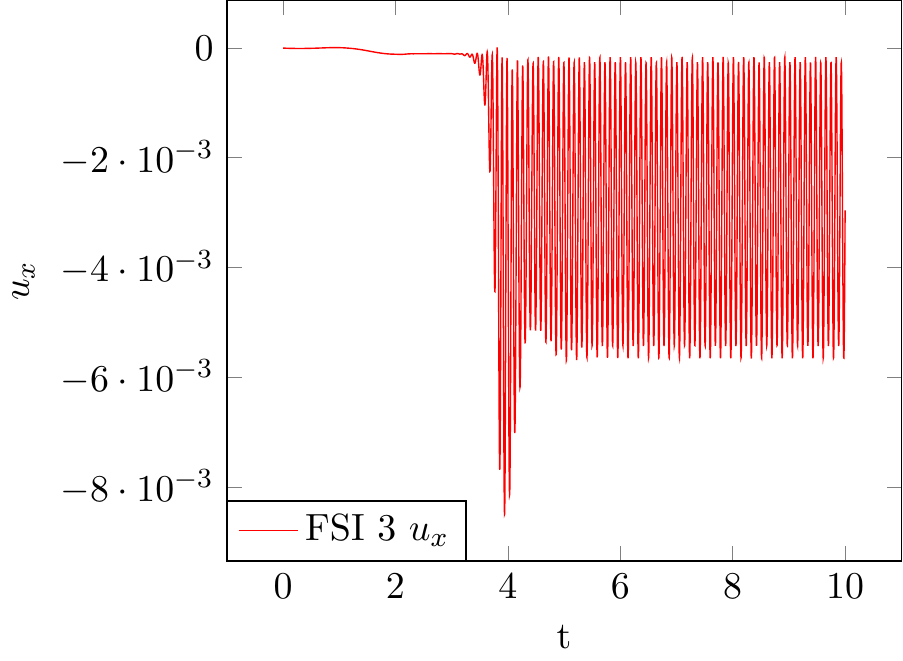}
\includegraphics[width=0.4\textwidth]{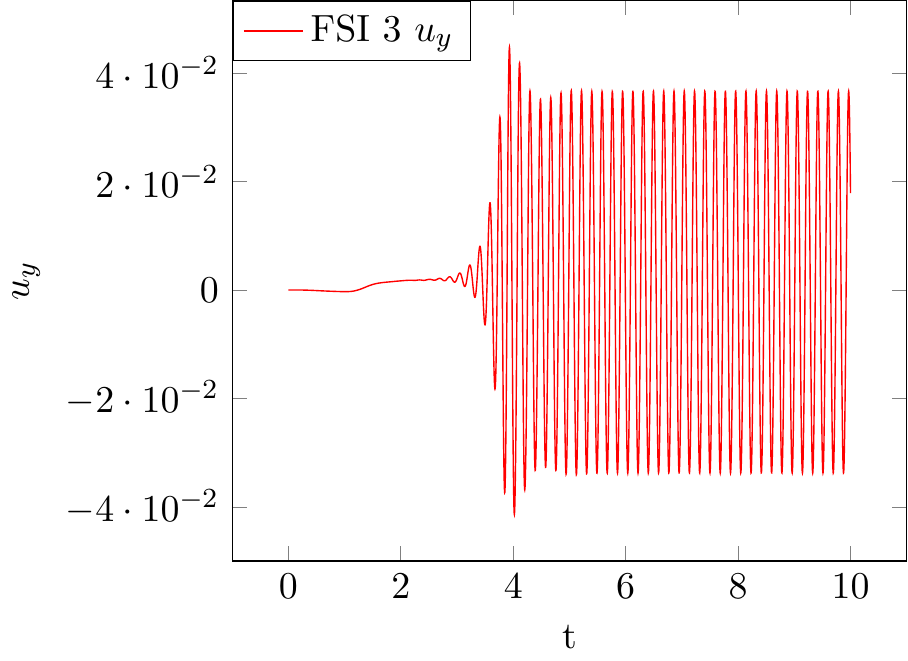}
\includegraphics[width=0.4\textwidth]{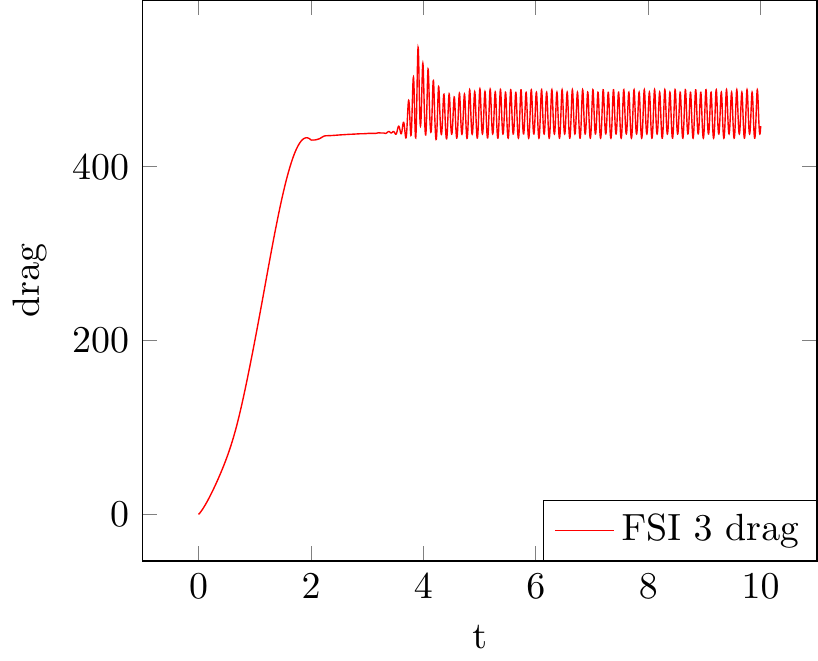}
\includegraphics[width=0.4\textwidth]{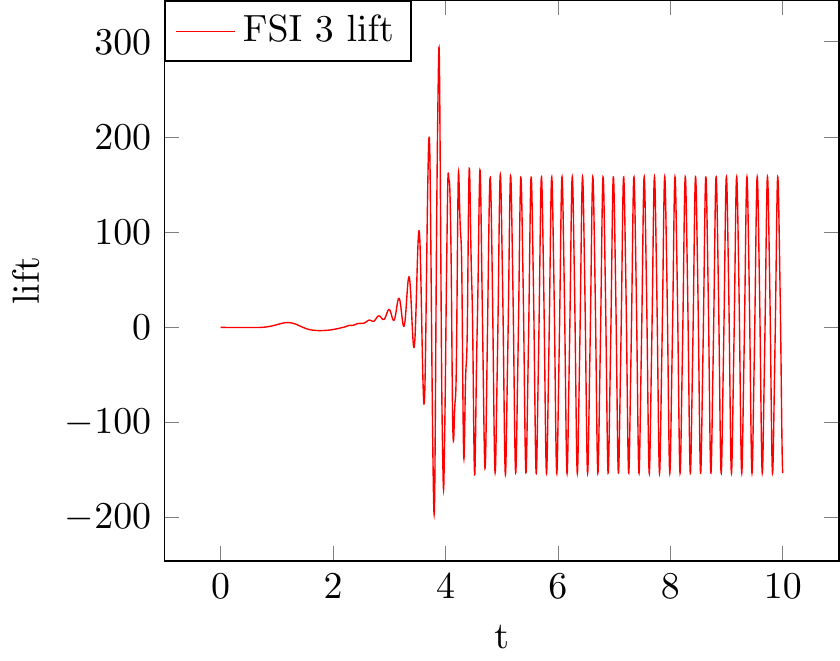}
\caption{Displacements $u_x$, $u_y$ at control point $A$ and drag, lift forces for FSI 3 benchmark.}
\label{fig:fsi3_plots}
\end{figure}

In the FSI 3 benchmark example the deformation does not become that critical as in FSI 2. Instead, due to the higher fluid velocity speed, the beam oscillates faster and thus, a smaller time step is needed. After time $t=5$ the solutions start to become periodically, see Figure \ref{fig:fsi3_plots}. The results are listed in Table \ref{tab:result_bl_hp_hdiv_fsi3}, where we observed a good match with \cite{THR10}.

\begin{table}
\centering
\input{fsi3_final_bl_minmax.out}
\caption{Results for FSI 3 with hp-refinement HDG and boundary layers around cylinder and flag.}
\label{tab:result_bl_hp_hdiv_fsi3}
\end{table}

\section*{Acknowledgements}
\label{sec:acknowledgments}
The authors acknowledge support from the Austrian Science Fund (FWF) through grant number W 1245.

\appendix
\section{Computation of derivatives of Piola transformation}
A straight forward calculation and the identity
\begin{align}
\partial_{\xref_i}\det(A)=\cof{A}:\partial_{\xref_i}A,
\end{align}
with $\cof{A}$ denoting the cofactor matrix of $A$ gives, with sum convention over $k$,
\begin{align}
\partial_{\xref_j}\left(\frac{1}{J}\Fmat u\right)_i &= \partial_{\xref_j}(\frac{1}{J}\Fmat_{ik}u_k)\nonumber\\
& =-\frac{1}{J^2}(\partial_{\xref_j}J)(\Fmat u)_i+\frac{1}{J}(\partial_{\xref_j}\Fmat_{ik}u_k+\Fmat_{ik}\partial_{\xref_j}u_k)\nonumber\\
& =-\frac{1}{J^2}\cof{\Fmat}:\partial_{\xref_j}\Fmat(\Fmat u)_i+\frac{1}{J}(\hesse^i_{jk}u_k+\Fmat_{ik}\partial_{\xref_j}u_k)\nonumber\\
& =-\frac{1}{J} \Fmat^{-T}:\partial_{\xref_j}\Fmat(\Fmat u)_i+\frac{1}{J}((\hesse^iu)_j+(\Fmat\gradref u)_{ij}),
\end{align}
where the identity $\cof{A} = \det(A)A^{-T}$ was used. 

For the second part the product rule and Schwarz's Theorem yields
\begin{align}
\frac{\partial}{\partial t}(\frac{1}{J}\Fmat u) &= -\frac{\partial_{t}J}{J^2}\Fmat u+\frac{1}{J}\gradref\dot{\varphi}u+\frac{1}{J}\Fmat\dot{u}\nonumber\\
& = \frac{1}{J}(\gradref\dot{\varphi}-\frac{J\tr{\gradref\dot{\varphi} \Fmat^{-1}}}{J}\Fmat)u+\frac{1}{J}\Fmat\dot{u}\nonumber\\
& =\frac{1}{J}(\gradref\dot{\varphi}-\tr{\gradref\dot{\varphi} \Fmat^{-1}}\Fmat)u+\frac{1}{J}\Fmat\dot{u},
\end{align}
where we exploited the following well known identities
\begin{align}
&\frac{\partial}{\partial t}\det(\Fmat)=J\Div[\xphys]{\dot{\varphi}}=J\tr{\gradref\dot{\varphi}\Fmat^{-1}}.
\end{align}

\bibliographystyle{acm}
\bibliography{cites}

\end{document}